\documentclass[12pt,oneside,a4paper]{article}
\pdfoutput=1

\usepackage{amsmath, amsthm, amsfonts, amssymb}
\usepackage{graphicx}
\usepackage{float}
\usepackage{color}
\usepackage{cite}
\usepackage[width=\textwidth,font=small,labelfont=bf,
labelsep=endash]{caption}
\usepackage{titlesec}

\titleformat{\section}
  {\normalfont\fontsize{13}{16}\bfseries}{\thesection}{1em}{}
\titleformat{\subsection}
  {\normalfont\fontsize{11}{14}\bfseries}{\thesubsection}{1em}{}

\normalsize
\long\def\@makecaption#1#2{%
  \vskip\abovecaptionskip
  \sbox\@tempboxa{\small #1. #2}%
  \ifdim \wd\@tempboxa >\hsize
    \small #1. #2\par
  \else
    \global \@minipagefalse
    \hb@xt@\hsize{\hfil\box\@tempboxa\hfil}%
  \fi
  \vskip\belowcaptionskip}

\newcommand\Lame {Lam\'e\ }

\newcommand\Schrodinger {Schr\"{o}dinger }

\newcommand\sign {\mathrm{sgn}}

\newcommand\csch {\mathrm{csch}}

\begin{document}

\title{\textbf{Elliptic String Solutions on $\mathbb{R}\times$S$^2$\\and Their Pohlmeyer Reduction}}
\author{Dimitrios Katsinis$^{1,2}$, Ioannis Mitsoulas$^3$ and Georgios Pastras$^2$}
\date{\small $^1$Department of Physics, National and Kapodistrian University of Athens,\\University Campus, Zografou, Athens 15784, Greece\\
$^2$NCSR ``Demokritos'', Institute of Nuclear and Particle Physics,\\Aghia Paraskevi 15310, Attiki, Greece\\
$^3$Department of Physics, School of Applied Mathematics and Physical Sciences,\\National Technical University, Athens 15780, Greece\linebreak \vspace{8pt}
\texttt{dkatsinis@phys.uoa.gr, mitsoula@central.ntua.gr, pastras@inp.demokritos.gr}}

\vskip .5cm

\maketitle

\abstract{We study classical string solutions on $\mathbb{R}^t \times $S$^2$ that correspond to elliptic solutions of the sine-Gordon equation. In this work, these solutions are systematically derived inverting Pohlmeyer reduction and classified with respect to their Pohlmeyer counterparts. These solutions include the  spiky strings and other well-known solutions, such as the BMN particle, the GKP string or the giant magnons, which arise as special limits, and reveal many interesting features of the AdS/CFT correspondence. A mapping of the physical properties of the string solutions to those of their Pohlmeyer counterparts is established. An interesting element of this mapping is the correspondence of the number of spikes of the string to the topological charge in the sine-Gordon theory. In the context of the sine-Gordon/Thirring duality, the latter is mapped to the Thirring model fermion number, leading to a natural classification of the solutions to fermionic objects and bosonic condensates. Finally, the convenient parametrization of the solutions, enforced by the inversion of the Pohlmeyer reduction, facilitates the study of the string dispersion relation. This leads to the identification of an infinite set of trajectories in the moduli space of solutions, where the dispersion relation can be expressed in a closed form by means of some algebraic operations, arbitrarily far from the infinite size limit.\linebreak \newline \textbf{Keywords:} Classical Strings, Integrable Systems, Pohlmeyer Reduction, AdS/CFT Correspondence}

\tableofcontents

\newpage

\setcounter{equation}{0}
\section{Introduction}
\label{sec:introduction}

Classical string solutions have played an important role in the understanding of the AdS/CFT correspondence \cite{ads_malda,ads_GKP,ads_witten}. According to the dictionary of the holographic duality, the dispersion relations of classical strings are related to the anomalous dimensions of gauge theory operators in the strong coupling limit. Matching the spectra on both sides of the holographic duality was a non-trivial quantitative test \cite{Frolov:2003qc,Beisert:2003xu,Frolov:2003xy,Beisert:2003ea,Roiban:2006jt} of the AdS/CFT correspondence and classical string solutions were necessary in order to perform such calculations. The standard methodology in the literature for this purpose, has been the use of an appropriate ansatz in order to reduce the classical string equations of motion and the Virasoro constraints to a system of equations for a set of unknown functions or parameters \cite{Arutyunov:2003uj,Arutyunov:2003za} (for a review see \cite{tseytlin_review}).

The matching of the spectra of the classical string in AdS$_5\times$S$^5$ and the $\mathcal{N} = 4$ SYM has also been studied with the help of methods from algebraic geometry. The sigma model \cite{Metsaev:1998it} of the Green-Schwarz superstring possesses a spectral curve, which is a manifestation of integrability \cite{Bena:2003wd}. On the field theory side, the anomalous dimensions of operators at strong coupling can be calculated using the Bethe ansatz \cite{Minahan:2002ve}. It has been shown that at specific limits, the spectra of the dual theories indeed match upon the identification of some parameters \cite{Kazakov:2004qf,Beisert:2005bm} (for a review see \cite{SchaferNameki:2010jy}). In this language, the classical string solutions are provided in terms of abstract hyperelliptic functions, that can be expressed in terms of conventional functions (algebraic or elliptic) only in the genus one case. Thus, although the problem of spectrum matching is formally understood, it is difficult to study and comprehend the generic structure.
 
A method for the construction of classical solutions in non-linear sigma models (NLSM) with a symmetric target space that is more systematic than the use of an arbitrary ansatz, but yet leads to solutions expressed in terms of functions with well understood properties, was initiated in \cite{bakas_pastras,Pastras:2016vqu}. In this approach, NLSM solutions are derived through the inversion of the Pohlmeyer reduction. Two-dimensional NLSMs with symmetric target spaces can be reduced to integrable systems, the so called symmetric space sine-Gordon systems (SSSG), which are multicomponent generalizations of the sine-Gordon equation. The older and most well-known example is the reduction of the O$(3)$ NSLM, which leads to the sine-Gordon equation \cite{Pohlmeyer:1975nb,Zakharov:1973pp}. The Pohlmeyer reduced system can always be derived from a local Lagrangian, which is a gauged Wess-Zumino-Witten model with an integrable potential \cite{Bakas:1993xh,Bakas:1995bm,FernandezPousa:1996hi,Miramontes:2008wt}. The Pohlmeyer reduction is equivalent to the Gauss-Codazzi equations for the embedding of the string worldsheet into the target space, which is in turn embedded into a flat enhanced space \cite{Lund:1976xd}. In this context, the fact that the target space is a symmetric space is directly connected to the integrability of the reduced model \cite{Eichenherr:1979ci,Eichenherr:1979hz}.

Even though it is straightforward to calculate the solution of the reduced theory that corresponds to a given solution of the original NLSM, the inversion of the Pohlmeyer reduction is a highly non-trivial process. This can be attributed to the non-local nature of the Pohlmeyer reduction, as well as to the fact that the mapping is many-to-one. Construction of NLSM solutions based on the inversion of the Pohlmeyer reduction has been performed in \cite{bakas_pastras} for strings propagating on AdS$_3$ and dS$_3$, and in \cite{Pastras:2016vqu} for minimal surfaces in H$^3$. These techniques can be applied for a particular class of solutions of the reduced system, which depend on a sole world-sheet coordinate. Given such a solution of the reduced system, the NLSM equations of motion become linear and solvable via separation of variables. Then, the geometric and Virasoro constraints are imposed and NLSM solutions are obtained. This procedure enables a systematic investigation of this class of NLSM solutions. In this work, we apply this method for strings that propagate on $\mathbb{R}^t \times $S$^2$. We argue that this study can be extended in a trivial manner to higher dimensional spheres.

String solutions belonging to this specific sector probe several interesting regimes of the spectrum of the AdS/CFT duality at specific limits. Berenstein, Maldacena and Nastase \cite{BMN} studied a particle moving at the equator of S$^5$ at the speed of light. Gubser, Klebanov and Polyakov \cite{GKP_string} studied a closed folded string that rotates around the north pole of the S$^2$ and its counter part, a string that is a rotating great circle. A few years later, Hofman and Maldacena \cite{Giant_Magnons} introduced the giant magnons. These are open strings, whose ends lie at the equator of the $S^2$ and move at the speed of light. They are the strong coupling, string theory counterpart of infinite size single-trace operators that contain one impurity. In \cite{single_spike_rs2,dual_spikes,multi,Dyonic_Giant_Magnons,helical,Kruczenski:2006pk} more general spiky string solutions are constructed. All these known solutions emerge naturally in our construction. We give a unified description and classification of all these string solutions in terms of their Pohlmeyer counterpart.

The paper is organized as follows. In section \ref{sec:Pohlmeyer}, we revisit the Pohlmeyer reduction of the NLSM describing strings propagating on $\mathbb{R}^t \times $S$^2$ that results in the sine-Gordon equation. In section \ref{sec:SG_elliptic}, we review the class of solutions of the sine-Gordon equation that can be expressed in terms of elliptic functions. In section \ref{sec:ell_strings}, it is shown that for these solutions of the sine-Gordon equation, the equations of motion of the NLSM separate into pairs of effective \Schrodinger problems. Each pair contains one flat potential, whereas the other one is the $n=1$ \Lame potential. We obtain the general solution for this system of equations and impose the appropriate constraints to effectively invert Pohlmeyer reduction. In section \ref{sec:properties}, we study various properties of the elliptic strings, with emphasis to the mapping of their properties to those of their Pohlmeyer counterparts. In section \ref{sec:dispersion}, we study the dispersion relations of the string solutions and finally, in section \ref{sec:discussion}, we discuss our results. Throughout the text, various properties of the Weierstrass elliptic and related functions are used. All the necessary formulae can be found in standard mathematical literature, e.g. \cite{Abramowitz}, or in the appendix of \cite{bakas_pastras}.

\setcounter{equation}{0}
\section{The Pohlmeyer Reduction of Strings Propagating on $\mathbb{R}^t \times $S$^2$}
\label{sec:Pohlmeyer}
In this section, we revisit the Pohlmeyer reduction of strings propagating on $\mathbb{R}^t \times $S$^2$ ($\mathbb{R}^t$ stands for the time dimension). The main difference of our approach to the original treatment \cite{Pohlmeyer:1975nb} is the implementation of a more general gauge, instead of the static gauge, which will facilitate the construction of the elliptic string solutions via the inversion of the Pohlmeyer reduction, in section \ref{sec:ell_strings}. This is the main reason we review the well-known Pohlmeyer reduction of strings propagating on the sphere here.

The basic ingredient of Pohlmeyer reduction is the embedding of the string world-sheet in a symmetric target space, which is in turn embedded in an enhanced higher-dimensional flat space. In the case of strings propagating on $\mathbb{R}^t \times $S$^2$, the higher dimensional flat space is $\mathbb{R}^{\left( 1 , 3 \right)}$. We denote the coordinates in the enhanced space as $X^0$, $X^1$, $X^2$ and $X^3$. Throughout this text, we use the following notation:
\begin{align}
A \cdot B &\equiv  - {A^0}{B^0} + {A^1}{B^1} + {A^2}{B^2} + {A^3}{B^3} ,\\
\vec A \cdot \vec B &\equiv {A^1}{B^1} + {A^2}{B^2} + {A^3}{B^3} .
\end{align}
Using this notation, the target space of the non-linear sigma model describing the propagation of strings on $\mathbb{R}^t \times $S$^2$ is simply the submanifold of the enhanced space:
\begin{equation}
\vec X \cdot \vec X = R^2 .
\label{eq:Pohlmeyer_geo_con}
\end{equation}
Writing the string action as a Polyakov action, we find,
\begin{equation}
S = T \int d\xi^+ d\xi^- \left( \left( \partial_+ X \right)\cdot \left( \partial_- X \right) + \lambda \left( \vec X \cdot \vec X - R^2 \right) \right) ,
\label{eq:Pohlmeyer_action}
\end{equation}
where $\xi^\pm$ are the right- and left-moving coordinates, $\xi^\pm \equiv \left( \xi^1 \pm \xi^0 \right) / 2$ and $T$ is the tension of the string.

The equations of motion that emerge from the action \eqref{eq:Pohlmeyer_action} read
\begin{align}
\partial_+ \partial_- X^0 &= 0 , \label{eq:Pohlmeyer_X0_eom}\\
\partial_+ \partial_- \vec X &= \lambda \vec X . \label{eq:Pohlmeyer_Xi_eom}
\end{align}
Obviously, the equation for the $X^0$ coordinate implies
\begin{equation}
X^0 = f_+ \left( \xi^+ \right) + f_- \left( \xi^- \right) .
\label{eq:Pohlmeyer_X0_solution}
\end{equation}
We may eliminate the Lagrange multiplier $\lambda$ from the equations of motion \eqref{eq:Pohlmeyer_Xi_eom}. The geometric constraint \eqref{eq:Pohlmeyer_geo_con} implies that $\partial_\pm\vec X \cdot \vec X = 0$. Upon another differentiation and the use of the equations of motion \eqref{eq:Pohlmeyer_Xi_eom}, we obtain
\begin{equation}
\lambda = - \frac{1}{R^2} \left( \partial_+ \vec X \right) \cdot \left( \partial_- \vec X \right) .
\end{equation}
Therefore, the equations of motion for the embedding functions $X^i$ assume the form
\begin{equation}
\partial_+ \partial_- \vec X = - \frac{1}{R^2} \left( \left( \partial_+ \vec X \right) \cdot \left( \partial_- \vec X \right) \right) \vec X .
\label{eq:Pohlmeyer_X_eom}
\end{equation}

The stress-energy tensor can be obtained by variation of the action with respect to the world-sheet metric. The off-diagonal components vanish identically, $T_{+-} = 0$, as a result of Weyl invariance. The diagonal elements equal
\begin{equation}
T_{\pm\pm} = \left( \partial_\pm X \right) \cdot \left( \partial_\pm X \right) .
\end{equation}
It follows that the Virasoro constraints assume the form, $
\left( \partial_\pm X \right) \cdot \left( \partial_\pm X \right) = 0$. Using the general solution for the embedding function $X^0$ given by equation \eqref{eq:Pohlmeyer_X0_solution}, the Virasoro constraints can be written as
\begin{equation}
\left( \partial_\pm \vec X \right) \cdot \left( \partial_\pm \vec X \right) = \left({f_\pm}' \right)^2 .
\label{eq:Pohlmeyer_Virasoro}
\end{equation}

The classical treatment of Pohlmeyer reduction takes advantage of the diffeomorphism invariance to set a specific form for the functions $f_\pm$, in particular selecting the static gauge, $X^0 = \mu \left( \xi_+ - \xi_- \right)$. For our purposes, it is more convenient to proceed without selecting a gauge and leave the advantage of this freedom for later use.

We define a basis in the enhanced three-dimensional space (the $\mathbb{R}^3$ subspace of $\mathbb{R}^{\left(1,3\right)}$),
\begin{equation}
\vec v_i = \left\{ \vec X , \partial_+ \vec X , \partial_- \vec X \right\} .
\end{equation}
The magnitude of the vectors $\vec v_i$ are fixed by the geometric and Virasoro constraints,
\begin{equation}
{\vec v}_1^2 = R^2 , \quad {\vec v}_2^2 = \left( {f_+}' \right)^2, \quad {\vec v}_3^2 = \left( {f_-}' \right)^2.
\label{eq:Pohlmeyer_basis_vectors_magnitudes}
\end{equation}
Furthermore, the geometric constraint upon differentiation yields $\partial_{\pm} \vec X \cdot \vec X = 0$ implying that ${\vec v}_1$ is perpendicular to ${\vec v}_2$ and ${\vec v}_3$,
\begin{equation}
{\vec v}_1 \cdot {\vec v}_2 = {\vec v}_1 \cdot {\vec v}_3 = 0 .
\label{eq:Pohlmeyer_basis_vectors_orthogonal}
\end{equation}

The only parameter that is not fixed by the constraints of the system is the angle between ${\vec v}_2$ and ${\vec v}_3$. We define it, as the Pohlmeyer field $\varphi$,
\begin{equation}
\left( \partial_+ \vec X \right) \cdot \left( \partial_- \vec X \right) := {f_+}' {f_-}' \cos \varphi.
\label{eq:Pohlmeyer_sG_field_definition}
\end{equation}

The relations \eqref{eq:Pohlmeyer_basis_vectors_magnitudes}, \eqref{eq:Pohlmeyer_basis_vectors_orthogonal} and \eqref{eq:Pohlmeyer_sG_field_definition} for the base vectors $\vec v_i$ can be used in order to decompose any vector $\vec V$ in the three-dimensional enhanced space in the base $\vec v _i$, as
\begin{multline}
\vec V = \frac{1}{R^2} \left( {\vec V \cdot {\vec v}_1} \right) {\vec v}_1 + \frac{{{f_ - }'\left( {\vec V \cdot {{\vec v}_2}} \right) - {f_ + }'\left( {\vec V \cdot {{\vec v}_3}} \right)\cos \varphi}}{{{{\left( {{f_ + }'} \right)}^2}{f_ - }'}}{{\vec v}_2} \\
+ \frac{{{f_ + }'\left( {\vec V \cdot {{\vec v}_3}} \right) - {f_ - }'\left( {\vec V \cdot {{\vec v}_2}} \right)\cos \varphi}}{{{{\left( {{f_ - }'} \right)}^2}{f_ + }'}}{{\vec v}_3} .
\label{eq:Pohlmeyer_decomposition}
\end{multline}
We decompose the derivatives of the base vectors into the base itself by introducing the $3\times3$ matrices $A^{\pm}$,
\begin{equation}
\partial_\pm {\vec v}_i = A^{\pm}_{ij} {\vec v}_j .
\end{equation}

By definition $\partial_+ {\vec v}_1 = {\vec v}_2 , \quad \partial_- {\vec v}_1 = {\vec v}_3 $, while the equations of motion imply that $\partial_+ {\vec v}_3 = \partial_- {\vec v}_2 = - {{f_+}' {f_-}'}/{R^2} \cos \varphi \, {\vec v}_1 $. So, the only basis vector derivatives left to calculate are $\partial_+ v_2 = \partial_+^2 X$ and $\partial_- v_3 = \partial_-^2 X$. The geometric constraint, upon two differentiations with respect to the same variable yields $( \partial _ \pm ^2\vec X ) \cdot \vec X =  - ( {{\partial _ \pm }\vec X} ) \cdot ( {{\partial _ \pm }\vec X} ) =  - {( {{f_ \pm }'} )^2} $. Furthermore, differentiating the Virasoro constraints, we get $( \partial^2_{\pm} \vec X ) \cdot ( \partial_{\pm} \vec X ) = {f_\pm}' {f_\pm}'' $. Finally, differentiation of the Pohlmeyer field definition \eqref{eq:Pohlmeyer_sG_field_definition} yields $( {\partial _ \pm ^2\vec X} ) \cdot ( {{\partial _ \mp }\vec X} ) = {f_ \pm }''{f_ \mp }'\cos \varphi - {f_ + }' {f_ - }' {\partial _ \pm } \varphi \sin \varphi $. Plugging the above into the decomposition formula \eqref{eq:Pohlmeyer_decomposition}, we get
\begin{align}
\partial_+ {\vec v}_2 &= - \frac{\left( {f_+}' \right)^2}{R^2} {\vec v}_1 + \left(\frac{{f_+}''}{{f_+}'} + \partial_+ \varphi \cot \varphi \right) {\vec v}_2 - \frac{{f_+}'}{{f_-}' \sin \varphi} {\vec v}_3 , \\
\partial_- {\vec v}_3 &= - \frac{\left( {f_-}' \right)^2}{R^2} {\vec v}_1 + \left(\frac{{f_-}''}{{f_-}'} + \partial_- \varphi \cot \varphi \right) {\vec v}_3 - \frac{{f_-}'}{{f_+}' \sin \varphi} {\vec v}_2 .
\end{align}
Putting everything together, the matrices $A^\pm$ assume the form,
\begin{align}
A^+ &= \begin{pmatrix}
0 & 1 & 0\\
- \frac{\left( {f_+}' \right)^2}{R^2} & \frac{{f_+}''}{{f_+}'} + \partial_+ \varphi \cot \varphi & -\frac{{f_+}'}{{f_-}' \sin \varphi}\\
- \frac{{f_+}' {f_-}'}{R^2} \cos a & 0 &0
\end{pmatrix} , \label{eq:Pohlmeyer_Aplus} \\
A^- &= \begin{pmatrix}
0 & 0 & 1\\
- \frac{{f_+}' {f_-}'}{R^2} \cos a & 0 &0\\
-\frac{\left( {f_-}' \right)^2}{R^2} & \frac{{f_-}''}{{f_-}'} + \partial_- \varphi \cot \varphi & - \frac{{f_-}'}{{f_+}' \sin \varphi}
\end{pmatrix} . \label{eq:Pohlmeyer_Aminus}
\end{align}
The matrices $A^\pm$ must obey the compatibility condition $\partial_+ \partial_- {\vec v}_i = \partial_- \partial_+ {\vec v}_i$, which can be written as the zero-curvature condition
\begin{equation}
\partial_- A^+ - \partial_+ A^- + \left[ {A^+ , A^-} \right] = 0 .
\end{equation}
Plugging the matrices \eqref{eq:Pohlmeyer_Aplus} and \eqref{eq:Pohlmeyer_Aminus} into the zero curvature condition yields
\begin{equation}
\partial_+ \partial_- \varphi = - \frac{{f_+}' {f_-}'}{R^2} \sin \varphi .
\label{eq:Pohlmeyer_pre_equation}
\end{equation}

The above equation can get simplified with the use of diffeomorphism invariance. We will not select the static gauge $f_\pm(\xi^\pm) := \pm \mu \xi^\pm$, but we will restrict ourselves to only what is necessary to put \eqref{eq:Pohlmeyer_pre_equation} to the form of the sine-Gordon equation, i.e. a more general ``linear'' gauge. We redefine the coordinates $\xi^\pm$, so that
\begin{equation}
f_\pm \left( \xi^\pm \right) := m_\pm \xi^\pm .
\end{equation}
The static and linear gauges are obviously connected via a worldsheet boost. In the following, we will construct classical string solutions, inverting the Pohlmeyer reduction, using the techniques of \cite{bakas_pastras}. The latter require solutions of the reduced system that depend solely on either $\xi^0$ or $\xi^1$. The freedom of the linear gauge selection allows the construction of classical string solutions, whose Pohlmeyer counterpart depends on a general linear combination of the worldsheet coordinates in the static gauge. Furthermore, it turns out that this freedom also facilitates the classification of the obtained solutions. Once the string solutions are found, one can always perform a boost to express them in the static gauge.

Calculating the induced metric on the world-sheet, using the Virasoro constraints \eqref{eq:Pohlmeyer_Virasoro} and the Pohlmeyer field definition \eqref{eq:Pohlmeyer_sG_field_definition}, we find
\begin{equation}
d{s^2} =  - {m_ + }{m_ - }{\sin ^2}\frac{\varphi }{2}\left( {{{\left( {d{\xi ^1}} \right)}^2} - {{\left( {d{\xi ^0}} \right)}^2}} \right) .
\end{equation}
Therefore, demanding that $\xi_0$ is the time-like parameter and $\xi_1$ is the space-like parameter sets $m_+ m_- < 0$. Then, the reduced system equation \eqref{eq:Pohlmeyer_pre_equation} assumes the form
\begin{equation}
\partial_+ \partial_- \varphi = \mu^2 \sin \varphi ,
\label{eq:Pohlmeyer_SGequation}
\end{equation}
where $\mu^2 := - {m_+} {m_-} / {R^2}$.

\setcounter{equation}{0}
\section{Elliptic Solutions of the Sine-Gordon Equation}
\label{sec:SG_elliptic}
In this section, we are going to find the solutions of the sine-Gordon equation \eqref{eq:Pohlmeyer_SGequation} that depend solely on one of the two world-sheet coordinates, i.e. they are either static or translationally invariant. In the following, the dot denotes differentiation with respect to $\xi^0$ and the prime denotes differentiation with respect to $\xi^1$.

Without loss of generality, we consider a solution that depends only on $\xi^0$, namely $\varphi \left( \xi^0 , \xi^1 \right) = \varphi_0 \left( \xi^0 \right)$. In this case, the sine-Gordon equation reduces to
\begin{equation}
{\ddot \varphi}_0 = - \mu^2 \sin \varphi_0 .
\end{equation}
This equation can be integrated once to yield
\begin{equation}
\frac{1}{2} {\dot \varphi}_0^2 - \mu^2 \cos \varphi_0 = E .
\label{eq:elliptic_energy_conservation}
\end{equation}

Similarly, had one considered static solutions that depend only on $\xi^1$, the only difference would be an overall sign. This sign can be absorbed defining $\varphi \left( \xi^0 , \xi^1 \right) = \pi + \varphi_1 \left( \xi^1 \right)$, which leads to
\begin{equation}
{\varphi_1}'' = - \mu^2 \sin \varphi_1 .
\end{equation}
It follows that static solutions can be produced by translationally invariant ones via an interchange of the coordinates and a shift of $\varphi$ by $\pi$.

Despite the simple symmetry that connects the translationally invariant solutions to the static ones, the two classes of solutions are characterized by qualitatively different Hamiltonian densities. The latter equals
\begin{equation}
\mathcal{H} = \frac{1}{2}{{\dot \varphi }^2} + \frac{1}{2}\varphi {'^2} - {\mu ^2}\cos \varphi .
\end{equation}
In the case of translationally invariant solutions, the Hamiltonian density is constant in both space and time and is equal to the integration constant $E$,
\begin{equation}
\mathcal{H} = E .
\end{equation}
On the contrary, in the case of static solutions, the Hamiltonian density is not constant, but it is a non-trivial function of $\xi^1$,
\begin{equation}
\mathcal{H} = \frac{1}{2}{\varphi _1}{'^2} - {\mu ^2}\cos {\varphi _1} = E - 2{\mu ^2}\cos {\varphi _1} = {\varphi _1}{'^2} - E .
\end{equation}
The momentum density is given by
\begin{equation}
\mathcal{P} =  - \varphi ' \dot \varphi 
\end{equation}
and it vanishes for both translationally invariant and static solutions.

It is clear that equation \eqref{eq:elliptic_energy_conservation} can be regarded as the conservation of energy of the simple pendulum. It is well known that the solutions to this problem can be expressed analytically in terms of elliptic functions. Indeed, performing the change of variable
\begin{equation}
2 y + \frac{E}{3} = - \mu^2 \cos \varphi_0 ,
\label{eq:elliptic_change_variable}
\end{equation}
the equation \eqref{eq:elliptic_energy_conservation} assumes the form
\begin{equation}
{y'}^2 = 4 y^3 - \left( \frac{E^2}{3} + \mu^4 \right) y - \frac{E}{3} \left( \left( \frac{E}{3} \right)^2 - \mu^4 \right) .
\label{eq:elliptic_SG_Weierstrass}
\end{equation}
This is the standard form of the Weierstrass equation ${y'}^2 = 4 y^3 - g_2 y - g_3 $, with specific values for the moduli equal to
\begin{equation}
g_2 = \frac{E^2}{3} + \mu^4 , \quad g_3 = \frac{E}{3} \left( \left( \frac{E}{3} \right)^2 - \mu^4 \right) .
\label{eq:elliptic_SG_moduli}
\end{equation}

\begin{figure}[ht]
\vspace{10pt}
\begin{center}
\begin{picture}(60,35)
\put(0,5){\includegraphics[width = 0.4\textwidth]{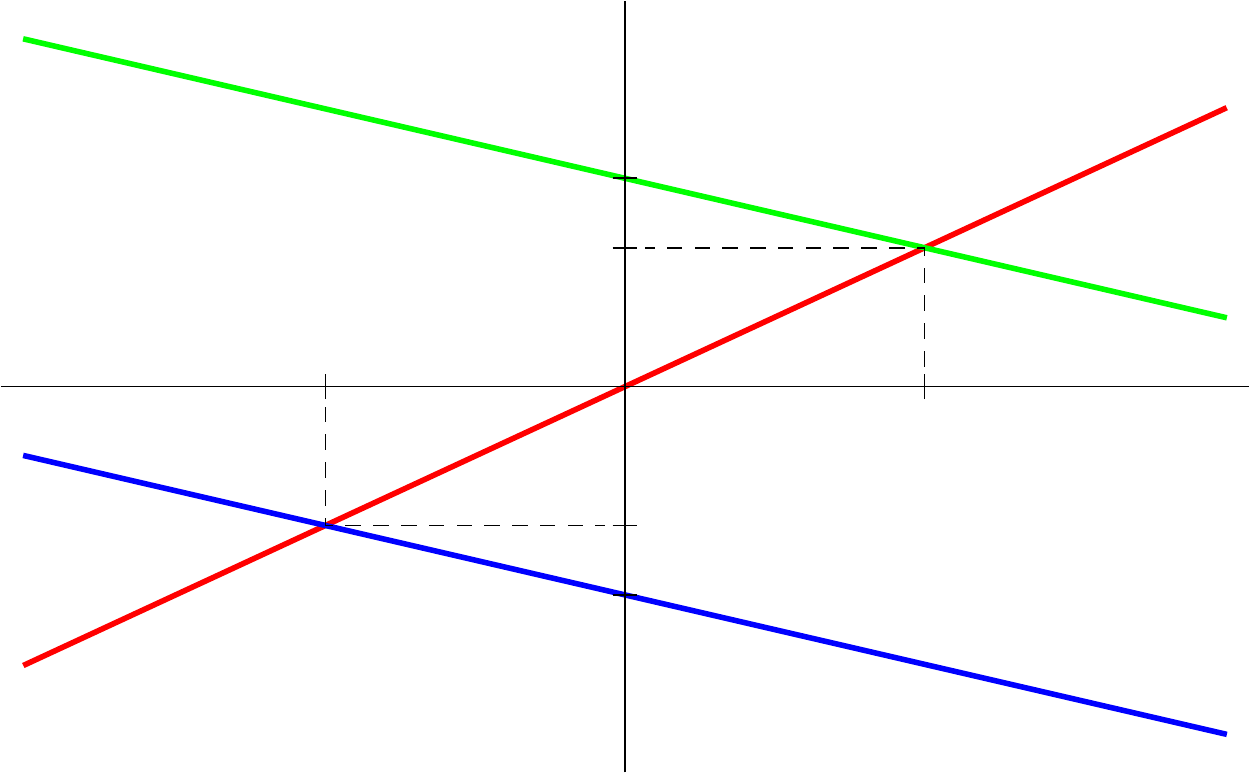}}
\put(44.5,9.5){\includegraphics[height = 0.15\textwidth]{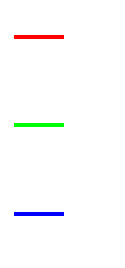}}
\put(7.25,18.75){$- \mu^2$}
\put(28.5,14.75){$\mu^2$}
\put(14,21){$\mu^2 / 3$}
\put(20.5,12.25){$-\mu^2 / 3$}
\put(49,22){$x_1$}
\put(49,17){$x_2$}
\put(49,12){$x_3$}
\put(19,31){$x_i$}
\put(40,16.5){$E$}
\put(44.5,10.5){\line(0,1){14}}
\put(44.5,10.5){\line(1,0){7.75}}
\put(52.25,10.5){\line(0,1){14}}
\put(44.5,24.5){\line(1,0){7.75}}
\end{picture}
\end{center}
\vspace{-5pt}
\caption{The roots of the cubic polynomial as function of the integration constant $E$}
\vspace{5pt}
\label{fig:roots}
\end{figure}
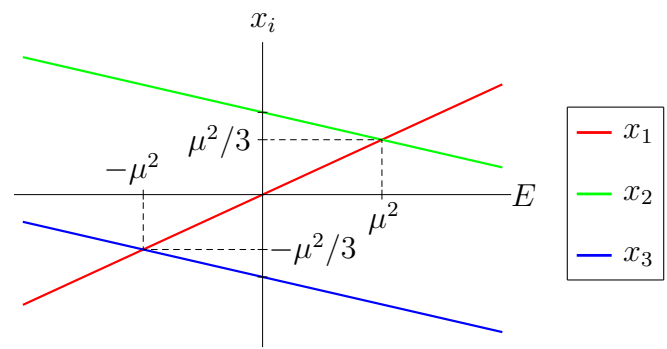

The general solution of the Weierstrass equation in the complex domain is provided by the Weierstrass elliptic function $\wp$. However, we are interested only in real solutions defined in the real domain. When the moduli $g_2$ and $g_3$ are real, the Weierstrass equation has one or two independent real solutions in the real domain, depending on the reality of the roots of the cubic polynomial $Q \left( y \right) = 4 y^3 - g_2 y - g_3 $. It turns out that the latter, with the moduli $g_2$ and $g_3$ given by \eqref{eq:elliptic_SG_moduli}, has always three real roots, namely,
\begin{equation}
x_1 = \frac{E}{3} , \quad x_2 = - \frac{E}{6} + \frac{\mu^2}{2} , \quad x_3 = - \frac{E}{6} - \frac{\mu^2}{2}.
\label{eq:elliptic_roots}
\end{equation}
The ordering of the three roots depends on the value of the integration constant $E$, as shown in figure \ref{fig:roots}.
Defining the ordered roots as $e_i$, where $e_1 > e_2 > e_3$, we have the identification between $x_i$ and $e_i$ that is shown in table \ref{tb:roots}.
\begin{table}[ht]
\centering
\begin{tabular}{|c |c| c|}
\hline
  & ordering of roots \\
\hline\hline
$E > \mu^2$ & $e_1 = x_1 , \quad e_2 = x_2 ,\quad e_3 = x_3$\\
\hline
$\left| E \right| < \mu^2$ & $e_1 = x_2 , \quad e_2 = x_1 ,\quad e_3 = x_3$\\
\hline
$E < - \mu^2$ & $e_1 = x_2 , \quad e_2 = x_3 , \quad e_3 = x_1$\\
\hline
\end{tabular}
\caption{The ordering of the roots}
\label{tb:roots}
\end{table}

When $Q \left( y \right)$ has three real roots, the fundamental periods of the Weierstrass elliptic function can be defined so that one of them is real and the other is purely imaginary. Let $2 \omega_1$ be the real one and $2 \omega_2$ be the imaginary one. Then, there are two distinct real solutions of the Weierstrass equation in the real domain, which read
\begin{align}
y &= \wp \left( x - x_0 \right) ,\\
y &= \wp \left( x - x_0 + \omega_2 \right) .
\end{align}
The first solution ranges between the largest of the roots and infinity, while the second one oscillates between the two smaller roots.

In order to acquire a real solution for $\varphi$, it is necessary that $y$ is real, but also it must satisfy
\begin{equation}
\left| 2 y + \frac{E}{3} \right| < \mu^2,
\label{eq:y_constraint}
\end{equation}
so that the change of variables \eqref{eq:elliptic_change_variable} maps a real $y$ to a real $\varphi$. The table \ref{tb:ranges} shows the range of $2 y + E / 3$ for each of the two solutions.
\begin{table}[ht]
\centering
\begin{tabular}{|c |c| c|}
\hline
 & range of $2 \wp ( x ) + E / 3$ & range of $2 \wp ( x + \omega_2 ) + E / 3$ \\
\hline\hline
$E > \mu^2$ & $2 y + E / 3 > E$ & $- \mu^2 < 2 y + E / 3 < \mu^2$\\
\hline
$\left| E \right| < \mu^2$ & $2 y + E / 3 > \mu^2$ & $- \mu^2 < 2 y + E / 3 < E$\\
\hline
$E < - \mu^2$ & $2 y + E / 3 > \mu^2$ & $E < 2 y + E / 3 < - \mu^2$\\
\hline
\end{tabular}
\caption{The range of $- \mu^2 \cos \varphi_0$ for both real solutions of the Weierstrass equation}
\label{tb:ranges}
\end{table}
It is clear that the unbounded solution does not correspond to a real solution for $\varphi$, as it does not satisfy the constraint \eqref{eq:y_constraint}. The bounded solution does correspond to a real solution for $\varphi$, as long as $E > - \mu^2$. This is expected from the physics of the simple pendulum. In all cases, the solution assumes the form
\begin{equation}
\cos \varphi_0 \left(\xi^0 ; E \right) = - \frac{1}{\mu^2}\left( 2 \wp \left( \xi^0 - \tau_0 + \omega_2 ; g_2 \left( E \right) , g_3 \left( E \right) \right) + \frac{E}{3} \right) .
\label{eq:elliptic_solution}
\end{equation}

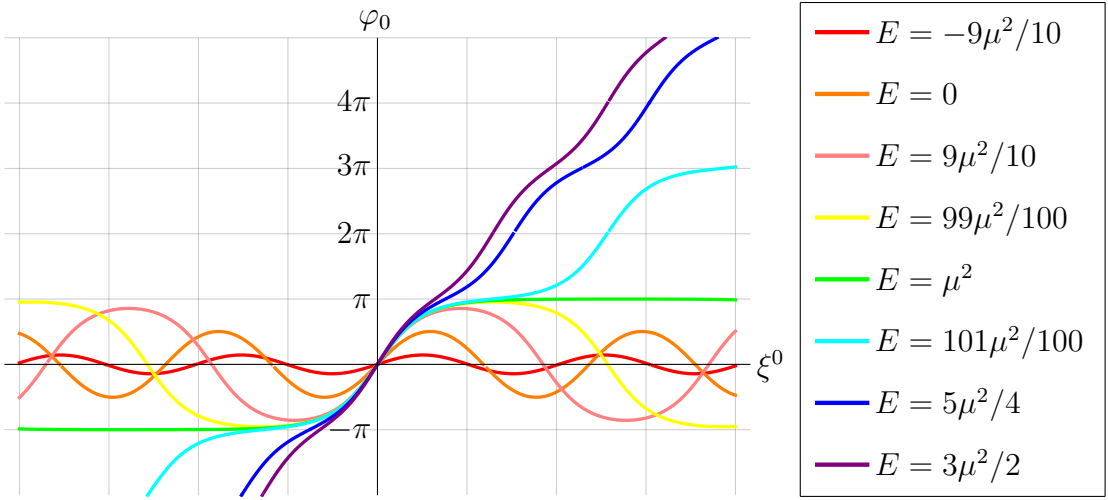
\begin{figure}[ht]
\vspace{10pt}
\begin{center}
\begin{picture}(90,41)
\put(0,1){\includegraphics[width = 0.6\textwidth]{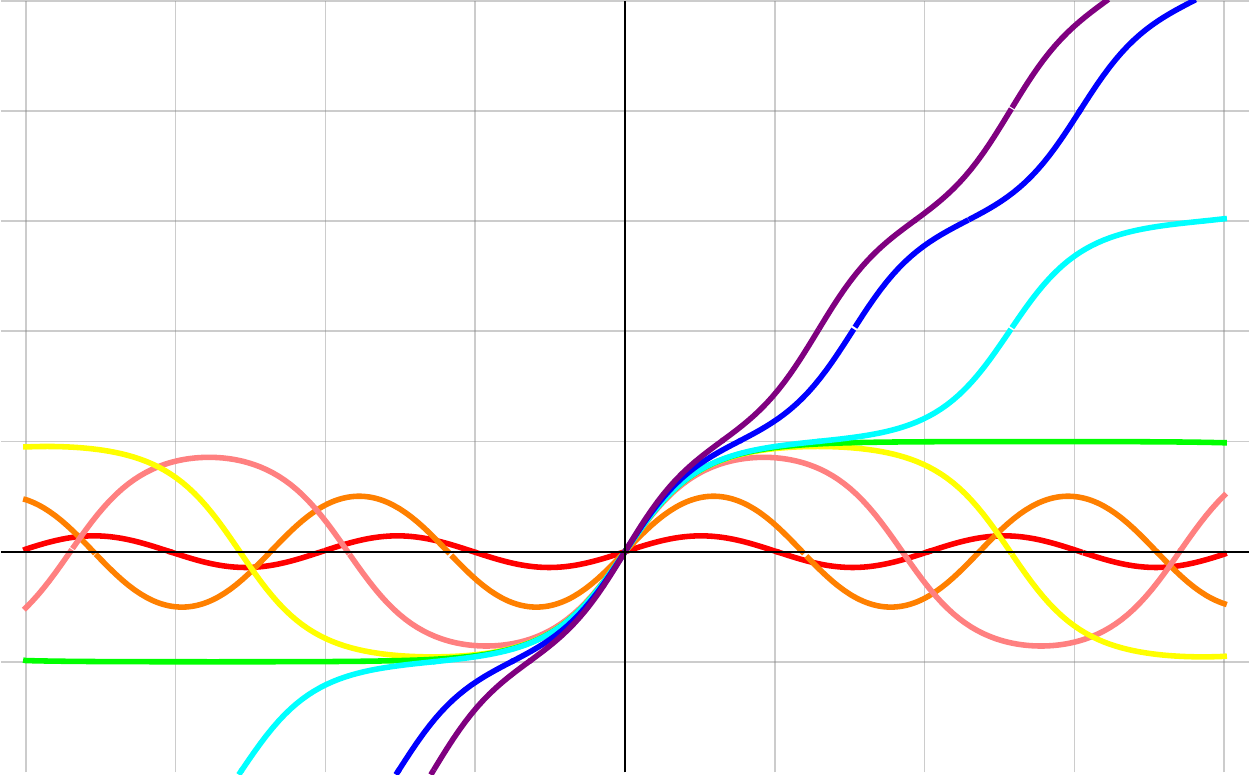}}
\put(28,16.25){$\pi$}
\put(26.75,21.5){$2\pi$}
\put(26.75,27){$3\pi$}
\put(26.75,32.25){$4\pi$}
\put(26,5.75){$-\pi$}
\put(64,1){\line(0,1){40}}
\put(64,1){\line(1,0){24.5}}
\put(88.5,1){\line(0,1){40}}
\put(64,41){\line(1,0){24.5}}
\put(64,0){\includegraphics[height = 0.4125\textwidth]{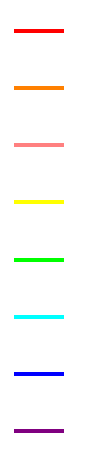}}
\put(70,37.75){$E=-9 \mu^2 /10$}
\put(70,32.75){$E=0$}
\put(70,27.75){$E=9 \mu^2 /10$}
\put(70,22.75){$E=99 \mu^2 /100$}
\put(70,17.75){$E= \mu^2 $}
\put(70,12.75){$E=101 \mu^2 /100$}
\put(70,7.75){$E=5 \mu^2 /4$}
\put(70,2.75){$E=3 \mu^2 /2$}
\put(28.5,39.25){$\varphi_0$}
\put(60.5,10.75){$\xi^0$}
\end{picture}
\end{center}
\vspace{-5pt}
\caption{The translationally invariant elliptic solutions of the sine-Gordon equation \eqref{eq:elliptic_solution_phi}, for various values of the energy constant $E$}
\vspace{5pt}
\label{fig:elliptic_solutions}
\end{figure}

Had one desired to find the solution for $\varphi_0$ itself, they would have to connect appropriate patches of $\varphi_0$, obeying equation \eqref{eq:elliptic_solution}, so that the solution is continuous and smooth. The appropriate combination of patches, which satisfies the initial conditions $\varphi_0 \left( \tau_0 \right) = 0$ and $\dot \varphi_0 \left( \tau_0 \right) = \sqrt {2\left( {E + {\mu ^2}} \right)} $, turns out to be
\begin{equation}
\varphi_0 \left(\xi^0 + \tau_0 \right) = \begin{cases}
{\left( { - 1} \right)^{\left\lfloor {\frac{\xi^0}{{2{\omega _1}}}} \right\rfloor }}\arccos \left[ - {\frac{1}{{{\mu ^2}}}\left( {2\wp \left( {\xi^0 + {\omega _2}} \right) + \frac{E}{3}} \right)} \right], & E < \mu^2 , \\
{\left( { - 1} \right)^{\left\lfloor {\frac{\xi^0}{{{\omega _1}}}} \right\rfloor }}\arccos \left[ - {\frac{1}{{{\mu ^2}}}\left( {2\wp \left( {\xi^0 + {\omega _2}} \right) + \frac{E}{3}} \right)} \right] + 2\pi \left\lfloor {\frac{{\xi^0} + {\omega _1}}{{2{\omega _1}}}} \right\rfloor , & E > \mu^2 ,
\end{cases}
\label{eq:elliptic_solution_phi}
\end{equation}
where $\arccos x$ is assumed to take values in $\left[ 0 , \pi \right]$. These solutions are plotted for various values of the energy constant $E$ in figure \ref{fig:elliptic_solutions}.

Similarly, the static elliptic solutions $\varphi_1 \left( \xi^1 \right)$ of the sine-Gordon equation, with boundary conditions $\varphi_0 \left( \sigma_0 \right) = \pi$ and ${\varphi_1} ' \left( \sigma_0 \right) = \sqrt {2\left( {E + {\mu ^2}} \right)} $, are
\begin{equation}
\varphi_1 \left(\xi^1 + \sigma_0 \right) = \begin{cases}
{\left( { - 1} \right)^{\left\lfloor {\frac{\xi^1}{{2{\omega _1}}}} \right\rfloor }}\arccos \left[ - {\frac{1}{{{\mu ^2}}}\left( {2\wp \left( {\xi^1 + {\omega _2}} \right) + \frac{E}{3}} \right)} \right] + \pi , & E < \mu^2 , \\
{\left( { - 1} \right)^{\left\lfloor {-\frac{\xi^1}{{{\omega _1}}}} \right\rfloor }}\arccos \left[ {\frac{1}{{{\mu ^2}}}\left( {2\wp \left( {\xi^1 + {\omega _2}} \right) + \frac{E}{3}} \right)} \right] + 2\pi \left\lfloor {\frac{{\xi^1} + {2{\omega _1}}}{{2{\omega _1}}}} \right\rfloor , & E > \mu^2 .
\end{cases}
\label{eq:elliptic_solution_phi_1}
\end{equation}

The solutions with $E < \mu^2$ are periodic, obeying $\varphi_{0/1} \left( \xi^{0/1} + 4 \omega_1 \right) = \varphi_{0/1} \left( \xi^{0/1} \right)$. We will call them the ``oscillatory'' solutions, inspired by the simple pendulum analogue of equation \eqref{eq:elliptic_energy_conservation}. The solutions with $E > \mu^2$ are quasi-periodic, obeying  $\varphi_{0/1} \left( \xi^{0/1} + 2 \omega_1 \right) = \varphi_{0/1} \left( \xi^{0/1} \right) + 2 \pi$. We will call them the ``rotating'' solutions.

\subsection{Double Root Limits}
When $E = \pm \mu^2$, two of the roots coincide, giving rise to some special limits of the elliptic solutions. In the case $E = - \mu^2$, the two smaller roots are both equal to $e_2 = e_3 = - \mu^2 /3$, and, thus, $\wp \left( {\xi^{0/1} + {\omega _2}} \right)$ tends to a constant equal to the double root. It follows that
\begin{align}
\varphi_0 \left(\xi^0 ; - \mu^2 \right) &= 0 ,\\
\varphi_1 \left(\xi^1 ; - \mu^2 \right) & = \pi .
\end{align}
Translationally invariant solutions tend to the stable vacuum of the sine-Gordon equation, whereas the static ones tend to the unstable vacuum.

For $E= \mu^2$, the two larger roots are both equal to $e_1 = e_2 = \mu^2 /3$. In this case the real period of the Weierstrass elliptic function diverges and the latter degenerates to a simply periodic hyperbolic function. It turns out that
\begin{align}
{\varphi _0}\left( {{\xi ^0};{\mu ^2}} \right) &= 4\arctan {e^{\mu \left( {{\xi ^0} - {\tau _0}} \right)}} + \pi , \\
{\varphi _1}\left( {{\xi ^1};{\mu ^2}} \right) &= 4\arctan {e^{\mu \left( {{\xi ^1} - {\sigma _0}} \right)}} .
\end{align}
The first one is an instanton solution evolving from the unstable vacuum $\varphi = - \pi$ to the unstable vacuum  $\varphi = + \pi$. The second one is the usual kink solution of the sine-Gordon equation, in the frame where it is static and localized in position $\xi_1 = \sigma_0$.

\setcounter{equation}{0}
\section{Elliptic String Solutions}
\label{sec:ell_strings}
\subsection{The Building Blocks of Elliptic Solutions}
\label{subsec:blocks}
Given a string configuration, it is a straightforward process to find the corresponding solution of the Pohlmeyer reduced system. The inverse problem is highly non-trivial due to the non-local nature of the Pohlmeyer reduction. This procedure comprises of using a given solution $\varphi$ of the reduced system and then solving the equations of motion
\begin{equation}
- \partial_0^2 \vec X + \partial_1^2 \vec X = \mu^2 \cos \varphi \vec X ,
\label{eq:blocks_eom}
\end{equation}
while simultaneously satisfying both the geometric 
\begin{equation}
\vec{X}\cdot\vec{X}=R^2
\label{eq:blocks_geometric}
\end{equation}
and the Virasoro constraints
\begin{equation}
\partial_\pm \vec{X}\cdot\partial_\pm \vec{X}=m_\pm^2 .
\label{eq:blocks_Virasoro}
\end{equation}
There is an advantage in finding a string solution starting from a given solution of the reduced system; the equations of motion have taken the form of the \emph{linear} differential equations \eqref{eq:blocks_eom}. Using a solution of the reduced system that depends on only one world-sheet coordinate provides an extra advantage; these linear differential equations are solvable using separation of variables \cite{bakas_pastras,Pastras:2016vqu},
\begin{equation}
X^i(\xi^0,\xi^1):=\Sigma^i(\xi^1) {\rm T}^i(\xi^0) .
\label{eq:blocks_separation}
\end{equation}

It is easy to show that in the case of a solution of the sine-Gordon equation that depends solely on $\xi^1$, the equations of motion \eqref{eq:blocks_eom} are written as pairs of effective \Schrodinger problems of the form,
\begin{align}
-{\Sigma^i}'' + \left( 2 \wp \left( \xi^1 + \omega_2 \right) + x_1 \right) \Sigma^i &= \kappa^i \Sigma^i , \label{eq:blocks_static_lame} \\
-{{\ddot {\rm T}}^i} &= \kappa^i {\rm T}^i . \label{eq:blocks_static_flat}
\end{align}
Similarly, in the case of solutions depending solely on $\xi^0$,
\begin{align}
-{\Sigma^i}'' &= \kappa^i \Sigma^i , \label{eq:blocks_trans_flat} \\
-{{\ddot {\rm T}}^i} + \left( 2 \wp \left( \xi^0 + \omega_2 \right) + x_1 \right) {\rm T}^i &= \kappa^i {\rm T}^i . \label{eq:blocks_trans_lame}
\end{align}

The form of the elliptic solutions of the sine-Gordon equation \eqref{eq:elliptic_solution} implies that in both cases, the non-trivial effective \Schrodinger problem \eqref{eq:blocks_static_lame} or \eqref{eq:blocks_trans_lame} assumes the form of the bounded $n = 1$ \Lame problem,
\begin{equation}
- \frac{d^2 y}{dx^2} + 2 \wp \left( x + \omega_2 \right) y = \lambda y .
\label{eq:blocks_lame_problem}
\end{equation}
The eigenfunctions of this problem are given by
\begin{equation}
y_\pm \left( x ; a \right) = \frac{\sigma \left( x + \omega_2 \pm a \right) \sigma \left( \omega_2 \right)}{\sigma \left( x + \omega_2 \right) \sigma \left( \omega_2 \pm a \right)} e^{- \zeta \left( \pm a \right) x} ,
\label{eq:blocks_lame_eigenfunctions}
\end{equation}
where the Weierstrass quasi-periodic functions $\zeta$ and $\sigma$ are defined as ${\zeta'}= - \wp$ and ${\sigma}' /{\sigma}=\zeta$.
The corresponding eigenvalue of both solutions $y_\pm$ is
\begin{equation}
\lambda = - \wp \left( a \right) .
\end{equation}

As long as $- \lambda$ is not equal to any of the roots, the pair of solutions \eqref{eq:blocks_lame_eigenfunctions} are linearly independent, and, thus, the general solution of \eqref{eq:blocks_lame_problem} can be written as a linear combination of the latter. At the limit $- \lambda$ becomes equal to any of the roots, both $y_\pm$ tend to
\begin{align}
y_\pm \left( x ; \omega_{2} \right) &= \sqrt{ \wp \left( x + \omega_2 \right) - e_3} ,\\
y_\pm \left( x ; \omega_{1,3} \right) &= \sqrt{e_{1,2} - \wp \left( x + \omega_2 \right)} .
\end{align}
In these cases, there is another linearly independent solution,
\begin{align}
\tilde y \left( x ; \omega_{2} \right) &= \sqrt{\wp \left( x + \omega_2 \right) - e_3} \left( \zeta \left( x + 2 \omega_2 \right) + e_3 x \right) , \\
\tilde y \left( x ; \omega_{1,3} \right) &= \sqrt{e_{1,2} - \wp \left( x + \omega_2 \right)} \left( \zeta \left( x + \omega_2 + \omega_{1,3} \right) + e_{1,2} x \right) .
\end{align}

When, the eigenvalue obeys $\lambda  < - e_1$ or $- e_2 < \lambda  < - e_3$, the eigenfunctions $y_\pm$ are real and they diverge exponentially at either plus or minus infinity. When the eigenvalue lies in the complementary segments, $\lambda  > - e_3$ or $- e_1 < \lambda  < - e_2$, the eigenfunctions $y_\pm$ are complex conjugate to each other and they are delta function normalizable Bloch waves.

Finally, the eigenfunctions $y_\pm$ obey the ``normalization'' relations
\begin{equation}
y_+ y_- = \frac{ \wp \left( x + \omega_2 \right) - \wp \left( a \right)}{e_3 - \wp \left( a \right)}
\label{eq:blocks_lame_yy}
\end{equation}
and
\begin{equation}
{y_+}' y_- - y_ + {y_-}' = -\frac{\wp' \left( a \right)}{e_3 - \wp \left( a \right)} . 
\label{eq:blocks_lame_yyprime}
\end{equation}

Summing up, there are three classes of solutions of the pair of effective \Schrodinger problems \eqref{eq:blocks_static_lame} and \eqref{eq:blocks_static_flat}, depending on the sign of the corresponding eigenvalue $\kappa_i$. Positive eigenvalues lead to embedding functions of the form
\begin{equation}
X = \left[ {c_ + ^1{y_ + }\left( {{\xi ^1};a} \right) + c_ - ^1{y_ - }\left( {{\xi ^1};a} \right)} \right]\cos \ell {\xi ^0} + \left[ {c_ + ^2{y_ + }\left( {{\xi ^1};a} \right) + c_ - ^2{y_ - }\left( {{\xi ^1};a} \right)} \right]\sin \ell {\xi ^0} ,
\label{eq:blocks_X_positive}
\end{equation}
where $\kappa = \ell^2 = - \wp \left( a \right) + x_1$. Negative eigenvalues lead to embedding functions
\begin{equation}
X = \left[ {c_ + ^1{y_ + }\left( {{\xi ^1};a} \right) + c_ - ^1{y_ - }\left( {{\xi ^1};a} \right)} \right]\cosh \ell {\xi ^0} + \left[ {c_ + ^2{y_ + }\left( {{\xi ^1};a} \right) + c_ - ^2{y_ - }\left( {{\xi ^1};a} \right)} \right]\sinh \ell {\xi ^0} ,
\label{eq:blocks_X_negative}
\end{equation}
where $\kappa = - \ell^2 = - \wp \left( a \right) + x_1$. Vanishing eigenvalue means that $\wp \left( a \right)$ equals to the root $x_1$, i.e. $a$ is one of the half-periods. Thus, the corresponding \Lame eigenfunctions degenerate to the form of eigenfunctions lying at the edge of the allowed bands. In general the solution is
\begin{equation}
X = \left[ {c_ + ^1y\left( {{\xi ^1};a} \right) + c_ - ^1\tilde y\left( {{\xi ^1};a} \right)} \right] + \left[ {c_ + ^2y\left( {{\xi ^1};a} \right) + c_ - ^2\tilde y\left( {{\xi ^1};a} \right)} \right]{\xi ^0} ,
\end{equation}
where $\wp \left( a \right) = x_1$. For ``normalization'' reasons that will become apparent later, we will consider only the part of this solution that can be taken as the limit of positive or negative eigenvalue solutions, i.e.
\begin{equation}
X = c \sqrt {{x_1} - \wp \left( {{\xi ^1} + {\omega _2}} \right)} .
\label{eq:blocks_X_zero}
\end{equation}

The embedding functions for the case of translationally invariant Pohlmeyer counterparts are identical to the above after an interchange of $\xi^0$ and $\xi^1$.

\subsection{Construction of Elliptic String Solutions}
\label{subsec:construction}

In section \ref{subsec:blocks}, we took advantage of the special form of the elliptic solutions of the sine-Gordon equation to solve the equations of motion via separation of variables. The general embedding function can then be written as a linear combination of the forms \eqref{eq:blocks_X_positive}, \eqref{eq:blocks_X_negative} and \eqref{eq:blocks_X_zero}. Then, in order to find a classical string solution, we need to find appropriate expressions for the three embedding functions $X^1$, $X^2$, and $X^3$ that satisfy the geometric constraint \eqref{eq:blocks_geometric} and the Virasoro constraints \eqref{eq:blocks_Virasoro}. The latter, expressed in terms of the coordinates $\xi^0$ and $\xi^1$, assume the form
\begin{align}
\left( {{\partial _0}\vec X} \right) \cdot \left( {{\partial _0}\vec X} \right) + \left( {{\partial _1}\vec X} \right) \cdot \left( {{\partial _1}\vec X} \right) &= \frac{{m_ + ^2 + m_ - ^2}}{2} , \\
2\left( {{\partial _0}\vec X} \right) \cdot \left( {{\partial _1}\vec X} \right) &= \frac{{m_ + ^2 - m_ - ^2}}{2} .
\end{align}
Since the embedding functions are solutions to the effective \Schrodinger problems \eqref{eq:blocks_static_lame} and \eqref{eq:blocks_static_flat}, we take advantage of the geometric constraint to write down the Virasoro constraints in the more handy form
\begin{align}
 - \left( {\partial _0^2\vec X} \right) \cdot \vec X - \left( {\partial _1^2\vec X} \right) \cdot \vec X &= \frac{{m_ + ^2 + m_ - ^2}}{2} , \label{eq:construction_Virasoro1}\\
 - 2\left( {{\partial _0}{\partial _1}\vec X} \right) \cdot \vec X &= \frac{{m_ + ^2 - m_ - ^2}}{2} . \label{eq:construction_Virasoro2}
\end{align}

In this work, we focus on the simplest choice, namely the use of a single eigenvalue for each component. The form of the geometric constraint enforces the two of the three components to correspond to the same positive eigenvalue and the third one to correspond to a vanishing one, i.e.
\begin{equation}
\vec X = \left( {\begin{array}{*{20}{c}}
{c_1^ + U_1^ + \left( {{\xi ^1};a} \right)\cos \ell {\xi ^0} + c_1^ - U_1^ - \left( {{\xi ^1};a} \right)\sin \ell {\xi ^0}}\\
{c_2^ + U_2^ + \left( {{\xi ^1};a} \right)\cos \ell {\xi ^0} + c_2^ - U_2^ - \left( {{\xi ^1};a} \right)\sin \ell {\xi ^0}}\\
{{c_3}\sqrt {{x_1} - \wp \left( {{\xi ^1} + {\omega _2}} \right)} }
\end{array}} \right) ,
\label{eq:construction_ansatz}
\end{equation}
where $\ell^2 = - \wp \left( a \right) + x_1$ and $U_{1,2}^\pm \left( {{\xi ^1};a} \right)$ are real linear combinations of $y_\pm \left( {{\xi ^1};a} \right)$.

Substituting the above into the geometric constraint \eqref{eq:blocks_geometric} and demanding that the terms proportional to $\sin \ell {\xi ^0} \cos \ell {\xi ^0}$, $\sin^2 \ell {\xi ^0}$ and $\cos^2 \ell {\xi ^0}$ vanish, yields 
\begin{align}
c_2^ +  =  - c_1^ - , &\quad c_2^ -  = c_1^ + ,\\
U_2^ +  = U_1^ - , &\quad U_2^ -  = U_1^ + .
\end{align}
Then, the geometric constraint assumes the form
\begin{equation}
{\left( {c_1^ + U_1^ + } \right)^2} + {\left( {c_1^ - U_1^ - } \right)^2} + c_3^2\left( {{x_1} - \wp \left( {{\xi ^1} + {\omega _2}} \right)} \right) = {R^2} .
\end{equation}
The normalization properties of the \Lame eigenfunctions \eqref{eq:blocks_lame_yy} imply that
\begin{align}
c_1^ +  = &c_1^ -  \equiv {c_1} ,\\
U_1^ +  = \frac{1}{2}\left( {{y_ + } + {y_ - }} \right),&\quad U_1^ -  = \frac{1}{{2i}}\left( {{y_ + } - {y_ - }} \right) .
\end{align}
It follows that in order to get a real solution, $y_\pm$ must be complex conjugate to each other, i.e. they must be Bloch wave eigenfunctions of the $n = 1$ \Lame problem. This constraints the parameter $\wp \left( a \right)$ to obey $\wp \left( a \right) < - e_3$, or $ - e_1 < \wp \left( a \right) < - e_2 $. Incorporating this into the geometric constraint, further simplifies it to the form
\begin{equation}
c_1^2{y_ + }{y_ - } + c_3^2\left( {{x_1} - \wp \left( {{\xi ^1} + {\omega _2}} \right)} \right) = {R^2} .
\end{equation}
The normalization property \eqref{eq:blocks_lame_yy} has an overall sign depending on whether the eigenstate belongs to the infinite ``conduction'' band $\wp \left( a \right) < - e_3$ or not. The only way that the $\xi^1$ dependence in the geometric constraint disappears is that $y_\pm$ are indeed such states, thus,
\begin{equation}
\wp \left( a \right) < - e_3 .
\end{equation}
This also implies that $a$ lies on the imaginary axis. Finally, absorbing the $e_3 - \wp \left( a \right)$ factor of \eqref{eq:blocks_lame_yy} into the definition of $y_\pm$, the geometric constraint reduces to
\begin{equation}
c_1 = c_3 \equiv c , \quad {c^2} = \frac{R^2}{x_1 - \wp \left( a \right)} = \frac{R^2}{\ell^2} .
\end{equation}

Taking the above into account, the ansatz \eqref{eq:construction_ansatz} assumes the form
\begin{equation}
\vec X = c\left( {\begin{array}{*{20}{c}}
{{\mathop{\rm Re}\nolimits} {y_ + }\left( {{\xi ^1};a} \right)\cos \ell {\xi ^0} + {\mathop{\rm Im}\nolimits} {y_ + }\left( {{\xi ^1};a} \right)\sin \ell {\xi ^0}}\\
{ - {\mathop{\rm Im}\nolimits} {y_ + }\left( {{\xi ^1};a} \right)\cos \ell {\xi ^0} + {\mathop{\rm Re}\nolimits} {y_ + }\left( {{\xi ^1};a} \right)\sin \ell {\xi ^0}}\\
{\sqrt {{x_1} - \wp \left( {{\xi ^1} + {\omega _2}} \right)} }
\end{array}} \right) .
\end{equation}
Substituting the above to the Virasoro constraint \eqref{eq:construction_Virasoro1} yields
\begin{equation}
{\ell ^2} = \frac{{m_ + ^2 + m_ - ^2}}{{4{R^2}}} + \frac{{3{x_1}}}{2} .
\end{equation}

Notice that the above equation implies that
\begin{equation}
- \wp \left( a \right) + {e_3} = {\left( {\frac{{{m_ + } + {m_ - }}}{{2R}}} \right)^2} > 0 ,
\end{equation}
as required in order for the \Lame eigenstates $y_\pm$ to lie in the infinite conduction band. The bound is saturated for ${m_ + } + {m_ - } = 0$. In this case, which corresponds to the special selection of the static gauge, the \Lame eigenfunctions $y_\pm$ are real and periodic functions that lie at the edge of the infinite conduction band. This limit is the equivalent to the GKP limit \cite{GKP_string}.

It is left to satisfy the Virasoro constraint \eqref{eq:construction_Virasoro2}. With the use of formula \eqref{eq:blocks_lame_yyprime}, the latter assumes the form
\begin{equation}
- i{c^2}\ell \wp '\left( a \right) = \frac{{m_ + ^2 - m_ - ^2}}{2} .
\label{eq:construction_Virasoro2_sign_of_a}
\end{equation}
The Weierstrass equation implies that
\begin{equation}
\begin{split}
 - {c^4}{\ell ^2}\wp {'^2}\left( a \right) &=  - 4{c^4}{\ell ^2}\left( {\wp \left( a \right) - {x_1}} \right)\left( {\wp \left( a \right) - {x_2}} \right)\left( {\wp \left( a \right) - {x_3}} \right)\\
 &= 4{c^4}{\ell ^4}\left( {{x_1} - {x_2} - {\ell ^2}} \right)\left( {{x_1} - {x_3} - {\ell ^2}} \right)\\
 &= 4{R^4}\left[ {{{\left( {{x_1} - \frac{{{x_2} + {x_3}}}{2} - {\ell ^2}} \right)}^2} - {{\left( {\frac{{{x_2} - {x_3}}}{2}} \right)}^2}} \right]\\
 &= 4{R^4}\left[ {{{\left( {\frac{{{x_1} + {x_2} + {x_3}}}{2} + \frac{{m_ + ^2 + m_ - ^2}}{{4{R^2}}}} \right)}^2} - {{\left( {\frac{{{\mu ^2}}}{2}} \right)}^2}} \right]\\
 &= {\left( {\frac{{m_ + ^2 - m_ - ^2}}{2}} \right)^2}
\end{split}
\end{equation}
and thus the Virasoro constraint \eqref{eq:construction_Virasoro2} is automatically satisfied without demanding further constraints in the free parameters of the solution. The subtlety in the sign can always by corrected by reflecting the parameter $a$, which corresponds to the transformation $y_\pm \to y_\mp$ or equivalently interchanging $m_+$ and $m_-$.

Putting everything together, the elliptic string solutions corresponding to static solutions of the sine-Gordon equation are written as
\begin{equation}
\vec X = \frac{R}{{\sqrt {{x_1} - \wp \left( a \right)} }}\left( {\begin{array}{*{20}{c}}
{{\mathop{\rm Re}\nolimits} \left({y_ + }\left( {{\xi ^1};a} \right){e^{ - i\ell {\xi ^0}}} \right)}\\
{ - {\mathop{\rm Im}\nolimits} \left({y_ + }\left( {{\xi ^1};a} \right){e^{ - i\ell {\xi ^0}}}\right)}\\
{{x_1} - \wp \left( {{\xi ^1} + {\omega _2}} \right)}
\end{array}} \right) .
\end{equation}

\setcounter{equation}{0}
\section{Properties of the Elliptic String Solutions}
\label{sec:properties}

In this section, we proceed to study the geometric characteristics of the string solutions derived in section \ref{sec:ell_strings} and their relation to the features of their Pohlmeyer counterparts. We indicate with index $0$, the elliptic string solutions that correspond to a translationally invariant solution of the sine-Gordon equation and with index $1$, the solutions with a static sine-Gordon counterpart. It turns out that the natural parametrization of our construction, which is based on the Weierstrass elliptic function, facilitates the study of the properties of the elliptic string solutions.

We take advantage of the fact that Bloch wave eigenfunctions of the \Lame potential are complex conjugates to each other and write them as
\begin{equation}
{y_ \pm }\left( {{\xi };a} \right) = \sqrt {\wp \left( {{\xi } + {\omega _2}} \right) - \wp \left( a \right)} {e^{ \pm i{\Phi }\left( {{\xi };a} \right)}} ,
\end{equation}
where
\begin{equation}
{\Phi }\left( {{\xi };a} \right) =  - \frac{i}{2}\ln \frac{{\sigma \left( {{\xi } + {\omega _2} + a} \right)\sigma \left( {{\omega _2} - a} \right)}}{{\sigma \left( {{\xi } + {\omega _2} - a} \right)\sigma \left( {{\omega _2} + a} \right)}} + i\zeta \left( a \right){\xi } .
\end{equation}
Notice that this function possesses the quasi-periodicity property
\begin{equation}
{\Phi }\left( {\xi  + 2{\omega _1};a} \right) = {\Phi }\left( {\xi ;a} \right) - 2i\left( {\zeta \left( {{\omega _1}} \right)a - \zeta \left( a \right){\omega _1}} \right) .
\label{eq:properties_quasiperiodicity_Phi}
\end{equation}
Thus, the elliptic string solutions assume the form
\begin{equation}
{\vec X}_{0/1} = \frac{R}{{\sqrt {{x_1} - \wp \left( a \right)} }}\left( {\begin{array}{*{20}{c}}
{\sqrt {\wp \left( {{\xi ^{0/1}} + {\omega _2}} \right) - \wp \left( a \right)} \cos\left( {\ell {\xi ^{1/0}} - {\Phi }\left( {{\xi ^{0/1}};a} \right)} \right)}\\
{\sqrt {\wp \left( {{\xi ^{0/1}} + {\omega _2}} \right) - \wp \left( a \right)} \sin \left( {\ell {\xi ^{1/0}} - {\Phi }\left( {{\xi ^{0/1}};a} \right)} \right)}\\
{\sqrt {{x_1} - \wp \left( {{\xi ^{0/1}} + {\omega _2}} \right)} }
\end{array}} \right) .
\end{equation}

Adopting spherical coordinates
\begin{align}
X^0 &= t ,\\
X^1 &= R \sin \theta \cos \varphi ,\\
X^2 &= R \sin \theta \sin \varphi ,\\
X^3 &= R \cos \theta ,
\end{align}
we acquire a parametric expression for the elliptic string solutions,
\begin{align}
t_{0/1} &= R\sqrt {{x_2} - \wp \left( a \right)} {\xi ^0} + R\sqrt {{x_3} - \wp \left( a \right)} {\xi ^1} , \label{eq:properties_t}\\
\cos \theta_{0/1}  &= \sqrt {\frac{{{x_1} - \wp \left( {{\xi ^{0/1}} + {\omega _2}} \right)}}{{{x_1} - \wp \left( a \right)}}} , \label{eq:properties_theta}\\
\varphi_{0/1}  &= - \sign ({\mathop{\rm Im}\nolimits} a) \sqrt {{x_1} - \wp \left( a \right)} {\xi ^{1/0}} - {\Phi }\left( {{\xi ^{0/1}};a} \right) . \label{eq:properties_phi}
\end{align}

Notice, that we have made the selection $m_+ + m_- > 0$ and $m_+ - m_- > 0$. The first choice is equivalent to the physical time $t$ being an increasing function of the time-like worldsheet coordinate $\xi^0$. Having selected one of the two above quantities to be negative, requires taking the opposite value of $a$ according to the Virasoro constraint \eqref{eq:construction_Virasoro2_sign_of_a}. We have restricted $a$ to take values in the segment of the imaginary axis with endpoints $\pm \omega_2$. Then, equation \eqref{eq:construction_Virasoro2_sign_of_a} implies that $\ell = - \sign ({\mathop{\rm Im}\nolimits} a) \sqrt{x_1 - \wp ( a )}$. From now on, for simplicity, we make the choice $\ell > 0$.

\subsection{Angular Velocity}
\label{subsubsec:omega}
Both classes of elliptic string solutions can be written in the form
\begin{equation}
f\left( {\theta ,\varphi  - \omega t} \right) = 0 .
\end{equation}
where
\begin{equation}
\omega_{0/1}  = \frac{\ell}{m_+ \pm m_-} , \quad \mathrm{or} \quad \left| \omega_{0/1} \right| = \frac{1}{R}\sqrt {\frac{{{x_1} - \wp \left( a \right)}}{{{x_{3/2}} - \wp \left( a \right)}}} .
\end{equation}
This angular velocity is a function of the gauge selection that we performed at the process of Pohlmeyer reduction.

Each class of elliptic string solutions is comprised of two subclasses, one corresponding to oscillating solutions of the sine-Gordon equation and one corresponding to rotating solutions of the latter. These are the well-known four classes of helical string solutions on the two-dimensional sphere \cite{helical} (see also \cite{single_spike_rs2,dual_spikes,multi,Dyonic_Giant_Magnons}). These two subclasses have some qualitative differences:
\begin{enumerate}
\item The solutions with rotating counterparts obey $x_1 > x_2$. Such solutions do not cross the equator; they lie between two circles, which are parallel to the equator and in the same semi-sphere. For example, in the case this is the north semi-sphere, these solutions obey
\begin{equation}
\theta_-  < \theta  < \theta_+ ,
\label{eq:ellliptic_properties_parallels_rotating}
\end{equation}
where
\begin{equation}
\theta_\pm = \arccos \sqrt {\frac{{{x_1} - {x_{2/3}}}}{{{x_1} - \wp \left( a \right)}}} .
\end{equation}
Both classes of such solutions are characterized by $\omega_{0/1} > 1 / R$. The angles $\theta_\pm$ that constrain the string on the sphere are also function of the gauge selection, since
\begin{equation}
\sin {\theta _ \mp } = \frac{1}{{R{\omega _{0/1}}}} .
\end{equation}
\item The solutions with oscillating counterparts obey $x_1 < x_2$. These solutions periodically cross the equator. They lie between two parallel circles, which are symmetrically placed above and below the equator, namely,
\begin{equation}
\theta_-  < \theta  < \pi  - \theta_- .
\label{eq:ellliptic_properties_parallels_oscillating}
\end{equation}
The angular velocity for solutions with static counterparts obeys $\omega_1 < 1 / R$. On the contrary, solutions with translationally invariant counterparts have $\omega_0 > 1 / R$. Notice that smoothness of the solution requires that $\cos \theta$ changes sign every time the string crosses the equator. Thus, the argument of the Weierstrass elliptic function should be altered by $4\omega_1$ in order to complete a whole period for $\theta$, in analogy to the period of the corresponding oscillating solutions of the sine-Gordon equation.
\end{enumerate}

In the static counterpart cases, the angular velocity tends to the critical value $\omega_{0/1} = 1 / R$, in the positive double root limit ($E \to \mu^2$), namely the limit of string solutions with a kink counterpart. The latter are the giant magnons \cite{Giant_Magnons}. In the translationally invariant counterpart cases, the angular velocity tends to the same critical value in the negative double root limit ($E \to - \mu^2$), namely the limit of string solutions corresponding to the stable vacuum of the sine-Gordon equation. This is the BMN particle solution \cite{BMN}.

Although, elliptic solutions with either static or translationally invariant counterparts accept a description of the form $f\left( {\theta ,\varphi  - \omega t} \right) = 0$, it is not clear whether this property should be conceived as a manifestation of rigid rotation or wave propagation. The fundamental difference between these two classes of solutions is that they can be written in a parametric form as
\begin{align}
\theta_{0/1}& = f \left( \xi^{0/1} \right) ,\\
\varphi_{0/1}  - \omega_{0/1} t_{0/1} &= g \left( \xi^{0/1} \right) .
\end{align}
In other words, $\theta$ and $\varphi - \omega t$ are parametrized in terms of the spacelike worldsheet coordinate in the static case. Thus, in this case, we may consider a given point of the string to be characterized by constant values of $\theta$ and $\varphi - \omega t$, implying rigidly rotating motion of the string. On the contrary, this is not the case for string solutions with translationally invariant counterparts, since in this case $\theta$ and $\varphi - \omega t$ are parametrized in terms of the timelike worldsheet coordinate. These solutions should be understood as wave propagation solutions.

\subsection{Periodicity Conditions}
\label{subsubsec:periodicity}
In order to better understand the form of the solutions, we may perform a worldsheet boost to convert to the static gauge,
\begin{align}
{\xi ^0} &= \gamma \left( {\sigma^0  - \beta \sigma^1 } \right) ,\\
{\xi ^1} &= \gamma \left( {\sigma^1  - \beta \sigma^0 } \right) ,
\end{align}
where
\begin{align}
\gamma \beta  &= \frac{1}{\mu }\sqrt {{x_3} - \wp \left( a \right)} ,\\
\gamma  &= \frac{1}{\mu }\sqrt {{x_2} - \wp \left( a \right)} .
\end{align}
Then, the elliptic string solutions assume the form
\begin{align}
t_{0/1} &= R \mu \sigma^0 , \label{eq:properties_static_t}\\
\cos \theta_{0/1}  &= \sqrt {\frac{{{x_1} - \wp \left( {{\gamma \left( {\sigma^{0/1}  - \beta \sigma^{1/0} } \right)} + {\omega _2}} \right)}}{{{x_1} - \wp \left( a \right)}}} , \label{eq:properties_static_theta}\\
\varphi_{0/1}  &= \sqrt {{x_1} - \wp \left( a \right)} {\gamma \left( {\sigma^{1/0}  - \beta \sigma^{0/1} } \right)} - {\Phi }\left( {{\gamma \left( {\sigma^{0/1}  - \beta \sigma^{1/0} } \right)};a} \right) . \label{eq:properties_static_phi}
\end{align}

\begin{figure}[ht]
\vspace{10pt}
\begin{center}
\begin{picture}(75,79.5)
\put(0,44.5){\includegraphics[width = 0.35\textwidth]{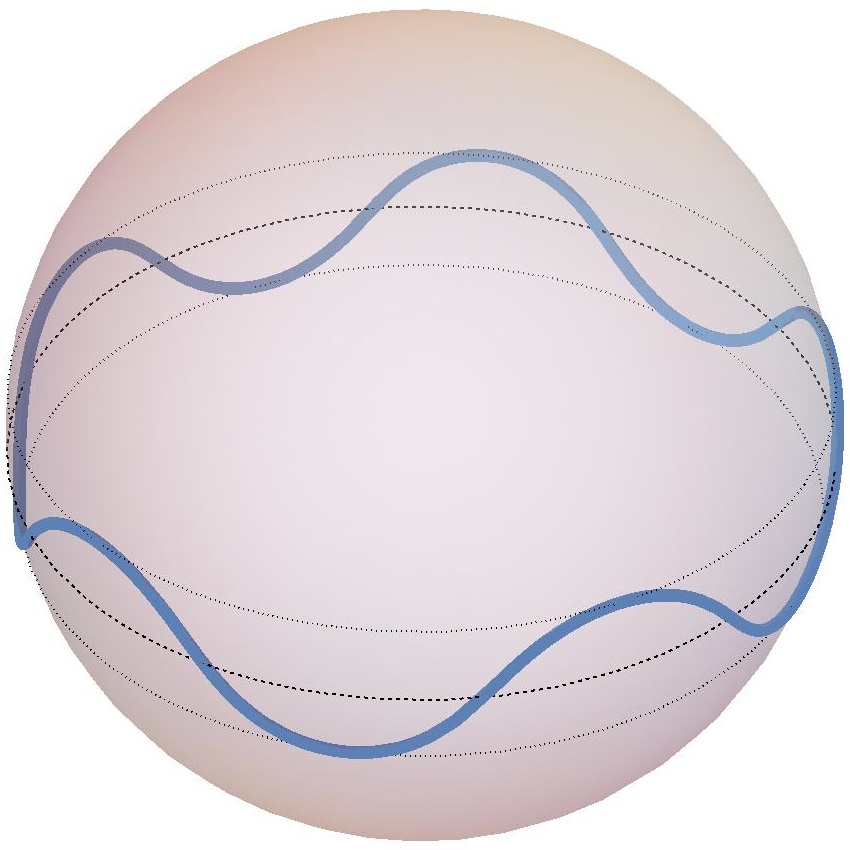}}
\put(40,44.5){\includegraphics[width = 0.35\textwidth]{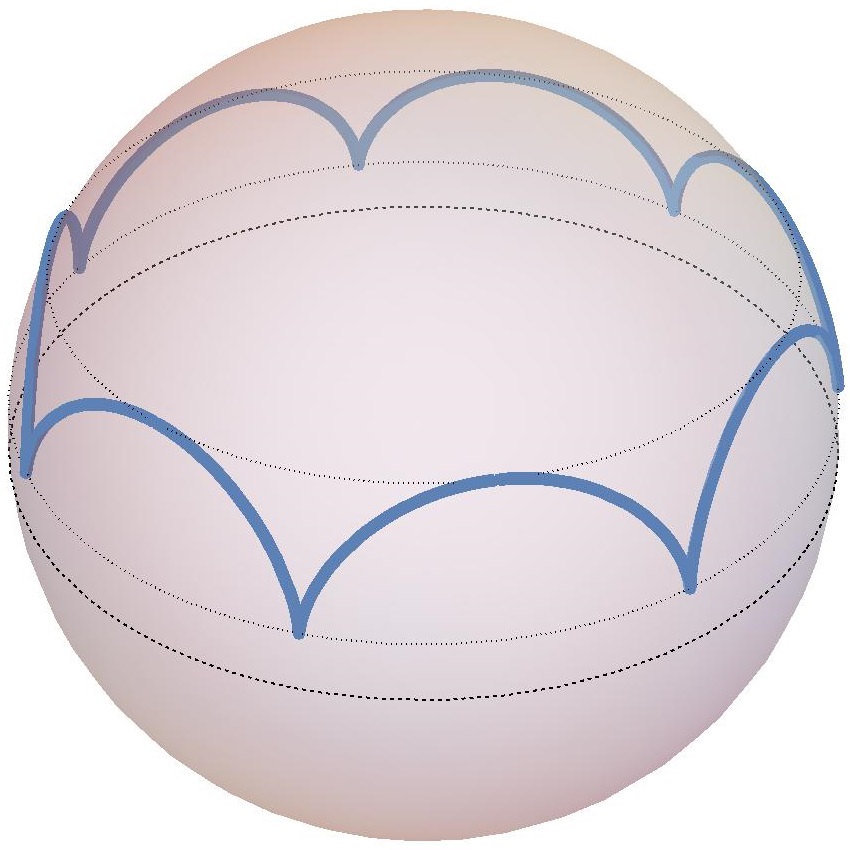}}
\put(0,4){\includegraphics[width = 0.35\textwidth]{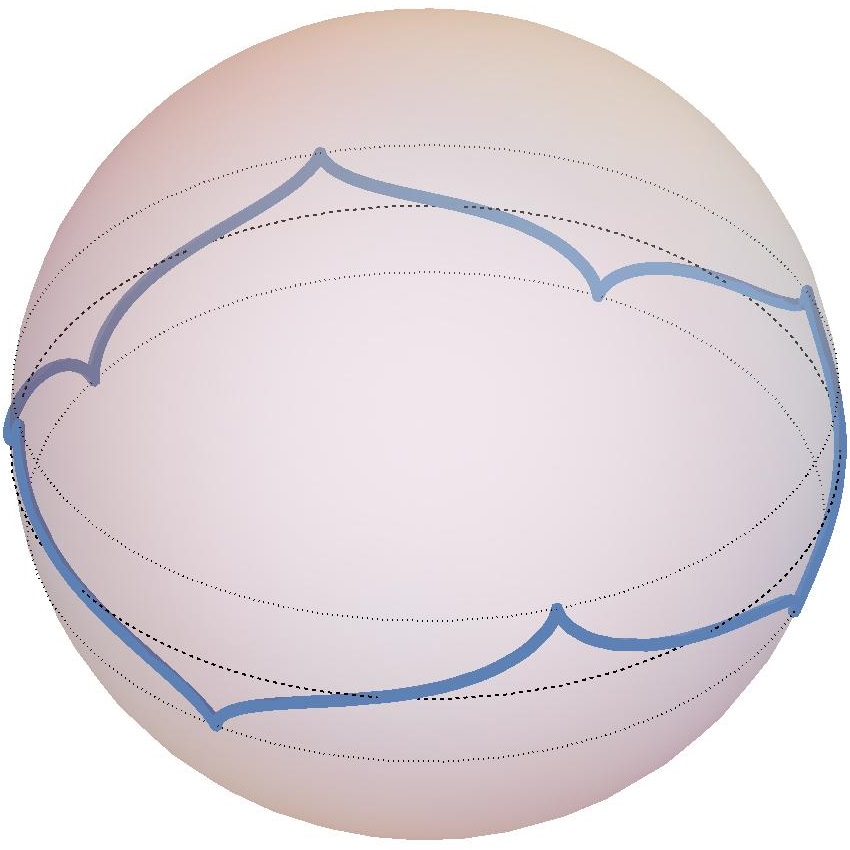}}
\put(40,4){\includegraphics[width = 0.35\textwidth]{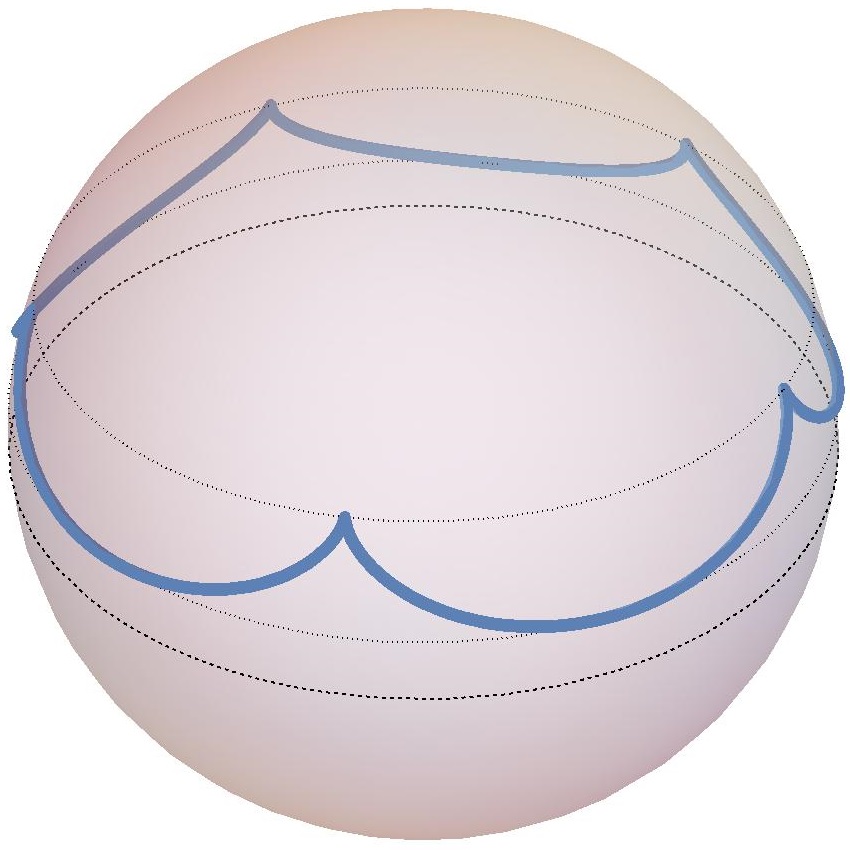}}
\put(9,42.75){static oscillating}
\put(11.5,40.5){counterpart}
\put(50,42.75){static rotating}
\put(51.5,40.5){counterpart}
\put(-0.5,2.25){translationally invariant oscillating}
\put(11.5,0){counterpart}
\put(39.5,2.25){translationally invariant rotating}
\put(51.5,0){counterpart}
\end{picture}
\end{center}
\vspace{-10pt}
\caption{The four classes of elliptic string solutions}
\vspace{5pt}
\label{fig:elliptic_strings}
\end{figure}
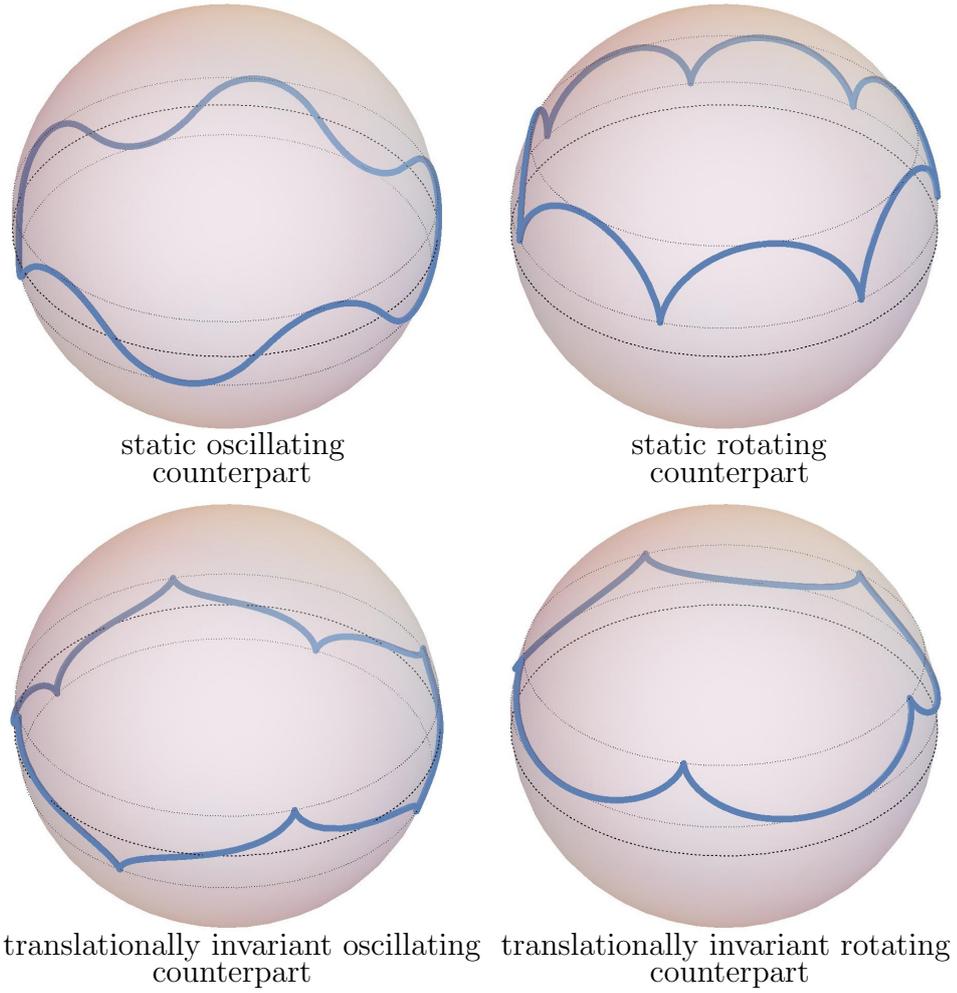

Equations \eqref{eq:properties_static_t}, \eqref{eq:properties_static_theta} and \eqref{eq:properties_static_phi} allow the visualization of a snapshot of the solution, as freezing the target space time $X^0$ is equivalent to freezing the worldsheet coordinate $\sigma^0$. The form of the four classes of elliptic string solutions defined in section \ref{subsubsec:omega} is depicted in figure \ref{fig:elliptic_strings}.

Clearly, equation \eqref{eq:properties_static_theta} implies that the angle $\theta$ is a periodic function of $\sigma^1$ in all cases. The period $\delta \sigma$ depends on the type of the solution. More specifically,
\begin{align}
\delta \sigma_0  &= \frac{{\delta \xi}}{\gamma \beta}, \\
\delta \sigma_1  &= \frac{{\delta \xi}}{\gamma },
\end{align}
where $\delta \xi$ is the real period/quasi-period of the corresponding sine-Gordon solution, namely
\begin{equation}
\delta \xi  = \begin{cases}
4{\omega _1}, &E < \mu^2 ,\\
2{\omega _1}, &E > \mu^2 .
\end{cases}
\label{eq:properties_real_sigma_period}
\end{equation}

Within a period $\delta \sigma$, the azimuthal coordinate $\varphi$ runs monotonically and its value changes by $\delta \varphi$, which is determined by the quasiperiodicity property \eqref{eq:properties_quasiperiodicity_Phi} of the function $\Phi$. It equals
\begin{equation}
\begin{split}
\delta \varphi_{0/1} &= \mp \delta \xi \left( {\frac{i}{{{\omega _1}}}\left( {\zeta \left( {{\omega _1}} \right)a - \zeta \left( a \right){\omega _1}} \right) + \frac{{\ell \left( {{m_ + } \mp {m_ - }} \right)}}{{{m_ + } \pm {m_ - }}}} \right) \\
&= \mp \delta \xi \left( {i\zeta \left( {{\omega _1}} \right)\frac{a}{\omega_1} - \left( {i\zeta \left( a \right) + \sqrt {\frac{{\left( {{x_1} - \wp \left( a \right)} \right)\left( {{x_{2/3}} - \wp \left( a \right)} \right)}}{{{x_{3/2}} - \wp \left( a \right)}}} } \right)} \right) \\
&= \mp i \delta \xi  \left( { \zeta \left( {{\omega _1}} \right)\frac{a}{\omega_1} + \zeta \left( {{\omega _{x_{3/2}}}} \right) - \zeta \left( a + {{\omega _{x_{3/2}}}} \right) } \right) ,
\end{split}
\end{equation}
where ${\omega _{x_i}}$ is the half-period corresponding to the root $x_i$, i.e. $\wp ( {\omega _{x_i}} ) = x_i$. The quantity $\delta \varphi_{0/1}$ has two contributions; one coming directly from the quasiperiodicity properties of the phase of the Bloch wave eigenfunctions of the $n=1$ \Lame potential and another one coming from the boost relating the static and linear gauges. Thus, the appropriate periodicity condition for closed elliptic string solutions without self-intersections is
\begin{equation}
i n_{0/1} \omega_1 \left( { \zeta \left( {{\omega _1}} \right)\frac{a}{\omega_1} + \zeta \left( {{\omega _{x_{3/2}}}} \right) - \zeta \left( a + {{\omega _{x_{3/2}}}} \right) } \right) = \pi ,
\end{equation}
where $n_{0/1}$ is an integer when $E>\mu^2$ and an even integer when $E<\mu^2$.


\subsection{Spikes}
\label{subsubsec:spikes}

In order to study the shape of the string, we differentiate the altitude $\theta$ and the azimuthal angle $\varphi$ with respect to the spacelike worldsheet variable $\sigma^1$. This yields
\begin{align}
\frac{{\partial {\theta _{0/1}}}}{{\partial {\sigma ^1}}} &=  \mp \frac{{\sqrt {{x_{3/2}} - \wp \left( a \right)} \wp '\left( {\gamma \left( {{\sigma ^{0/1}} - \beta {\sigma ^{1/0}}} \right) + {\omega _2}} \right) }}{{2\mu \sqrt {{x_1} - \wp \left( a \right)} \sqrt {\wp \left( {\gamma \left( {{\sigma ^{0/1}} - \beta {\sigma ^{1/0}}} \right) + {\omega _2}} \right) - \wp \left( a \right)} }} ,\\
\frac{{\partial \varphi_{0/1} }}{{\partial \sigma ^1}} &= \mp \frac{{\sqrt {\left( {{x_1} - \wp \left( a \right)} \right)\left( {{x_{2/3}} - \wp \left( a \right)} \right)} \left( {{x_{3/2}} - \wp \left( {\gamma \left( {\sigma^{0/1}  - \beta \sigma^{1/0} } \right) + {\omega _2}} \right)} \right) }}{{\mu \left( {\wp \left( {\gamma \left( {\sigma^{0/1}  - \beta \sigma^{1/0} } \right) + {\omega _2}} \right) - \wp \left( a \right)} \right)}} .
\end{align}
For solutions with static counterparts, ${{\partial \varphi_{0/1} }}/{{\partial \sigma ^1}}$ vanishes only if $x_2$ is equal to $e_2$, i.e. only for rotating solutions of the sine-Gordon equation. In this case, it vanishes when $\gamma \left( \sigma^1 - \beta \sigma^0 \right) = \left( 2 n + 1 \right) \omega_1$, where $n \in \mathbb{Z}$. For solutions with translationally invariant counterparts, it vanishes when $\gamma \left( \sigma^1 - \beta \sigma^0 \right) = 2 n \omega_1$, where $n \in \mathbb{Z}$. The locations where ${{\partial \varphi_{0/1} }}/{{\partial \sigma ^1}}$ vanishes are lying at altitude 
\begin{equation}
\sin \theta_{0/1}^{\rm{spike}}  = \sin \theta_\mp . 
\end{equation}
Therefore, in such locations the altitude $\theta$ acquires an extremal value implying that its derivative changes sign. Indeed, ${{\partial {\theta _{0/1}}}}/{{\partial {\sigma ^1}}}$ also vanishes at these positions. At this points, $\partial \theta / \partial \varphi$ diverges as
\begin{equation}
\left| {\frac{{\partial \theta }}{{\partial \varphi }}} \right| \sim \left| {\frac{{\wp '\left( {\gamma \left( {{\sigma ^{0/1}} - \beta {\sigma ^{1/0}}} \right) + {\omega _2}} \right)}}{{\wp \left( {\gamma \left( {{\sigma ^{0/1}} - \beta {\sigma ^{1/0}}} \right) + {\omega _2}} \right) - {x_{3/2}}}}} \right| \sim \frac{1}{{\sqrt {\left| {\wp \left( {\gamma \left( {{\sigma ^{0/1}} - \beta {\sigma ^{1/0}}} \right) + {\omega _2}} \right) - {x_{3/2}}} \right|} }} .
\end{equation}
It follows that these positions are positions of spikes.

The Pohlmeyer field in the position of a spike assumes the value
\begin{equation}
\varphi_{\mathrm{Pohlmeyer}}^{\mathrm{spike}} = 2 n \pi , \quad n \in \mathbb{Z} .
\end{equation}
This justifies the form of the elliptic string solutions presented in figure \ref{fig:elliptic_strings}. Translationally invariant oscillatory solutions of the sine-Gordon equation oscillate around $\varphi = 0$. For this reason, the corresponding strings have spikes that appear periodically. Half of those spikes point towards the north pole of the sphere and half of them towards the south pole, corresponding to the Pohlmeyer field being equal to zero with positive or negative derivative. On the contrary, static oscillatory solutions of the sine-Gordon equation oscillate around $\varphi = \pi$ and as a result, the corresponding strings do not have spikes. Both classes of rotating solutions of the sine-Gordon equation are always increasing (or decreasing) functions and therefore periodically cross positions with $\varphi = 2 n \pi$ with the same derivative. For this reason, the string solutions with rotating counterparts present spikes periodically, which point to the same pole of the sphere.

It is easy to show that
\begin{equation}
R\omega_{0/1} \sin \theta_{0/1}^{\rm{spike}} = 1 ,
\end{equation}
i.e. the spikes are moving at the speed of light. In the static counterpart case, the spike may have the interpretation of a given point of the string, which due to initial conditions, is moving at the speed of light and therefore cannot change velocity no matter what forces are exerted on it. In the translationally invariant counterpart case, which has the interpretation of wave propagation, a given point of the string is spiky at a given time instant, when this point reaches the speed of light, as a result of the propagation of a wave pattern along the string, and gets violently reflected. Since the elliptic strings preserve their shape as time evolves, spikes cannot get in contact, in order to study their interactions. It would be interesting to study the outcome of the collision of such spiky points; this requires the investigation of string solutions with more complicated Pohlmeyer counterparts.

The fact that spikes appear at locations where the Pohlmeyer field is a multiple of $2\pi$ is not a coincidence. Writing down the Virasoro constraints in the static gauge yields
\begin{align}
{\left| {{\partial _0}\vec X} \right|^2} &= {R^2}{\mu ^2}{\cos ^2}\frac{\varphi }{2} ,\\
{\left| {{\partial _1}\vec X} \right|^2} &= {R^2}{\mu ^2}{\sin ^2}\frac{\varphi }{2} .
\end{align}
Thus, any singular point of the string, i.e. a spike, which necessarily is characterized by vanishing $ {{\partial _1}\vec X} $, is a point where the Pohlmeyer field is a multiple of $2\pi$. Furthermore, the Virasoro constraints imply that these points have $\left| {{\partial _0}\vec X} \right| = R \mu$, which combined to the fact that at the static gauge $t = R \mu \sigma^0$ implies that the spikes move at the speed of light. Notice that the Virasoro constraints do not imply that any point of the string where the Pohlmeyer field is a multiple of $2 \pi$, is necessarily a singular spiky point. However, the latter is also true in the class of elliptic string solutions.

\subsection{Topological Charge and the Sine-Gordon/Thirring Duality}
The limit of the elliptic solutions of the sine-Gordon equation at plus and minus spatial infinity is well-defined only in the vacuum and kink limits. Therefore, a topological charge can be naturally defined only in these cases. However, in the case of string configurations with appropriate periodicity conditions, the Pohlmeyer field obeys periodic and not asymptotic conditions, namely, 
\begin{equation}
\varphi \left( \sigma^0 , \sigma^1 + \delta \sigma \right) - \varphi \left( \sigma^0 , \sigma^1 \right) = 2 n \pi , \quad n \in \mathbb{Z} .
\end{equation}
Therefore, a topological charge in the Pohlmeyer reduced theory can be defined in such solutions, which obviously equals $n$. We have seen that a spike appears whenever the Pohlmeyer field assumes a value that is an integer multiple of $2 \pi$. It follows that
\begin{equation}
n = \mathrm{number} \; \mathrm{of}\; \mathrm{spikes}.
\end{equation}
Notice that spikes pointing to opposite poles of the sphere have opposite contributions to this conserved charge, i.e. they function as spikes and ``anti-spikes''. This is evident in the case of string solutions with translationally invariant oscillating counterparts (see figure \ref{fig:elliptic_strings}). Conservation of the topological charge in the Pohlmeyer reduced theory implies some kind of ``conservation of the number of spikes'', which should also apply in more complicated string solutions, where spikes may get in touch and interact.

It is well known that the sine-Gordon equation is S-dual to the Thirring model \cite{Coleman:1974bu}. The Lagrangian densities of the two theories are
\begin{align}
\mathcal{L}_{\mathrm{SG}} &= \frac{1}{2}{\partial _\mu }\varphi {\partial ^\mu }\varphi  + \frac{{{\alpha _0}}}{{{\beta ^2}}}\cos \beta \varphi ,\\
\mathcal{L}_{\mathrm{Th}} &= i\bar \Psi {\gamma ^\mu }{\partial _\mu }\Psi  - {m_0}\bar \Psi \Psi  - \frac{g}{2}\left( {\bar \Psi {\gamma ^\mu }\Psi } \right)\left( {\bar \Psi {\gamma _\mu }\Psi } \right) .
\end{align}
The Thirring model possesses a global symmetry, namely
\begin{equation}
\Psi \to e^{i a} \Psi .
\end{equation}
This gives rise to a conserved current
\begin{equation}
j^\mu = \bar \Psi {\gamma ^\mu }\Psi 
\end{equation}
and a conserved charge, namely the fermion number,
\begin{equation}
N = \int {d{\sigma ^1}\bar \Psi {\gamma ^0 }\Psi } .
\end{equation}

The duality implies that the parameters and fields of the two dual theories are connected as,
\begin{align}
\frac{{4\pi }}{{{\beta ^2}}} &= 1 + \frac{g}{\pi } ,\\
 - \frac{\beta }{{2\pi }}{\varepsilon ^{\mu \nu }}{\partial _\nu }\varphi  &= \bar \Psi {\gamma ^\mu }\Psi , \\
\frac{{{\alpha _0}}}{{{\beta ^2}}}\cos \beta \varphi  &=  - {m_0}\bar \Psi \Psi .
\end{align}
The classical limit corresponds to $\beta = 1$ \cite{Kimel:1976vz}. According to the above, the conserved current of the Thirring model can be expressed in terms of the sine-Gordon field as
\begin{align}
j^0 &= - \frac{1}{2\pi} \partial_1 \varphi , \\
j^1 &= \frac{1}{2\pi} \partial_0 \varphi ,
\end{align}
and, thus, the fermion number assumes the form
\begin{equation}
N = - \frac{1}{{2\pi }}\int {d{\sigma ^1}{\partial _1}\varphi } = - n ,
\end{equation}
which equals the opposite of the topological charge in the Pohlmeyer reduced theory, and, thus, the number of spikes.

The above correspondence naively implies that in the picture of the Thirring model, the string solutions with rotating counterparts can be considered as multi-fermion states. On the contrary, solutions with oscillating Pohlmeyer counterparts have the natural interpretation of bosonic condensates. However, notice that the sine-Gordon/Thirring duality is a full quantum weak to strong duality. Thus, the above statement should be viewed cautiously, since taking the classical limit of a strongly coupled quantum theory is in general non-trivial.

It would be interesting to investigate this duality in the framework of string theory. Type IIB superstring theory in AdS$_n\times$S$^n$ is self-S-dual, with the closed strings being S-dual to D1-branes \cite{Hull:1994ys,Witten:1995ex}. This hints that the spiky elliptic strings should be S-dual to D1-brane configurations, whose Pohlmeyer counterpart has non-trivial fermion number equal to the number of spikes of the original string solutions. The investigation of this correspondence requires the derivation of elliptic superstring solutions propagating on the full AdS$_n\times$S$^n$ space and their parallel study in the corresponding supersymmetric Pohlmeyer reduced theory.

\subsection{Interesting Limits and the Moduli Space of Solutions}
The elliptic string solutions have some very well known special limits, which are very simple to study in our parametrization. We do so for the completeness of our presentation. At these limits, two of the three roots $x_1$, $x_2$ and $x_3$ coincide, and, thus, the Weierstrass elliptic function degenerates to a simply periodic function, either trigonometric or hyperbolic. There are two such cases:

In the limit $E \to -\mu^2$, the two negative roots coincide and the solutions reduce to
\begin{align}
\cos \theta_{0/1}  &= 0,\\
\varphi_{0/1}  &=  - \mu \sigma^{0/1} .
\end{align}
being a hoop around the equator \cite{Roiban:2006jt} in the static counterpart case and the BMN particle \cite{BMN} travelling along the equator at the speed of light in the translationally invariant counterpart case. Notice that in this limit, the string worldsheet degenerates to a one-dimensional manifold. This is not unexpected, since in this limit, the solution of the Pohlmeyer reduced system degenerates to the vacuum solution of the sine-Gordon equation, meaning that the vectors $\partial_+ X$ and $\partial_- X$ become parallel. This property is present to other NLSMs (e.g. see \cite{Pastras:2016yaa}).

Similarly, in the limit $E \to \mu^2$, the two positive roots coincide and the solutions degenerate to
\begin{align}
\cos \theta_{0/1}  &= i\sinh \left( {\mu a} \right){\mathop{\rm sech}\nolimits} \left[ {i\mu \left( {\csch\left( {\mu a} \right)\sigma^{0/1}  - \coth \left( {\mu a} \right)\sigma^{1/0} } \right)} \right] , \\
\varphi_{0/1}  &= \mu \sigma^{1/0}  + \frac{i}{2}\ln \frac{{\cosh \left[ {i\mu \left( {\csch\left( {\mu a} \right)\sigma^{0/1}  - \coth \left( {\mu a} \right)\sigma^{1/0} - ia} \right)} \right]}}{{\cosh \left[ {i\mu \left( {\csch\left( {\mu a} \right)\sigma^{0/1}  - \coth \left( {\mu a} \right)\sigma^{1/0}  + ia} \right)} \right]}} ,
\end{align}
being the giant magnon \cite{Giant_Magnons} with angular opening $\delta \varphi = 2 i \mu a$ in the case of solutions with static counterparts and the single spike \cite{single_spike_rs2} in the case of solutions with translationally invariant counterparts.

The above two limits are specific values for the integration constant $E$. For a given value of this constant, the parameter $a$ may take any value on the imaginary axis on the linear segment defined by the origin and the half-period $\omega_2$. Another interesting limit is the special selection $a = - \omega_2$ or $\wp \left( a \right) = x_3$. This is the case where the linear gauge coincides with the static gauge. Had we restricted Pohlmeyer reduction to the static gauge, the method applied in section \ref{sec:ell_strings} for the construction of the elliptic string solutions would have resulted to these special solutions only. In this limit, the solution assumes the form
\begin{align}
\cos \theta_{0/1} &= \sqrt \frac{{x_1} - \wp \left( {\sigma^{0/1}  + {\omega _2}} \right)}{x_1 - x_3} ,\\
\varphi_{0/1} &= \sqrt {{x_1} - {x_3}} \sigma^{1/0} .
\end{align}
In the case of static oscillating counterparts, this is a great circle crossing the two poles and rotating with angular velocity $\omega_1$, whereas in the case of a static rotating counterpart this is an arc of a great circle centered at one of the two poles and rotating with angular velocity $\omega_1$ so that its endpoints have the speed of light. This is the well known GKP string solution \cite{GKP_string}. Notice that this limit is always compatible with the periodicity conditions corresponding to the value $n_1=2$.

In the case of solutions with translationally invariant counterparts, the above solution describes a hoop being always parallel to the equator which shrinks to a point at the pole of the sphere and then extends again. In the case of oscillating solutions it extends further than the equator and then shrinks again to the opposite pole before it starts re-extending, whereas in the case of rotating solutions it extends up to a maximum size and then it shrinks again to the same pole. These solutions, although they have a translationally invariant Pohlmeyer counterpart are spikeless. This is due to the coincidence of the static gauge to the linear one. As there is no need for a worldsheet boost to convert to the static gauge, the singular behaviour characterizes solely the time evolution of the string and not its shape. These solutions satisfy the periodicity conditions with $n_0 = 0$. The coordinate $\sigma^1$ takes values in $\left[ {0 , 2 \pi / \sqrt{ x_1 - x_3}} \, \right)$ to complete one hoop.

The opposite limit to the above is $a / \omega_2 \to 0$. In this limit the solution assumes the form
\begin{align}
\cos {\theta _{0/1}} &= \left| a \right|\sqrt {{x_1} - \wp \left( {\frac{{{\sigma ^{0/1}} - {\sigma ^{1/0}}}}{{\mu \left| a \right|}} + {\omega _2}} \right)} ,\\
{\varphi _{0/1}} &= \frac{{{x_2}}}{\mu }{\sigma ^{0/1}} - \frac{{{x_3}}}{\mu }{\sigma ^{1/0}} + \left| a \right|\left( {\zeta \left( {\frac{1}{{\mu \left| a \right|}}\left( {{\sigma ^{0/1}} - {\sigma ^{1/0}}} \right) + {\omega _2}} \right) - \zeta \left( {{\omega _2}} \right)} \right) .
\end{align}
This describes strings that have the shape of the general solution, lying very close to the equator and being characterized by a small angular opening $\delta \varphi$. In this limit, the static gauge and the linear one are connected via a boost by a velocity close to the speed of light. These string solutions are the ``speeding strings'' limit \cite{Mikhailov:2003gq}.

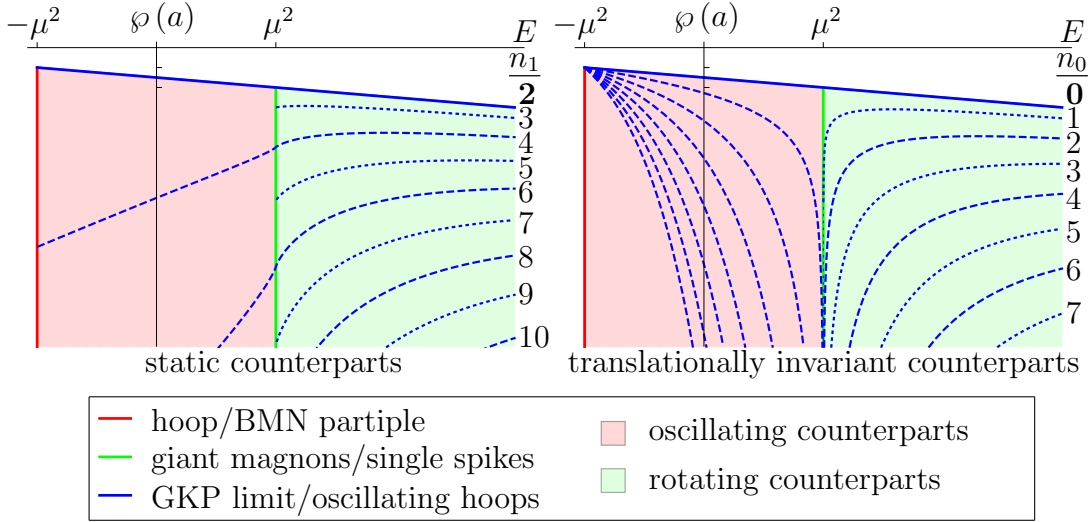
\begin{figure}[ht]
\vspace{10pt}
\begin{center}
\begin{picture}(88,42.5)
\put(1,14){\includegraphics[width = 0.4\textwidth]{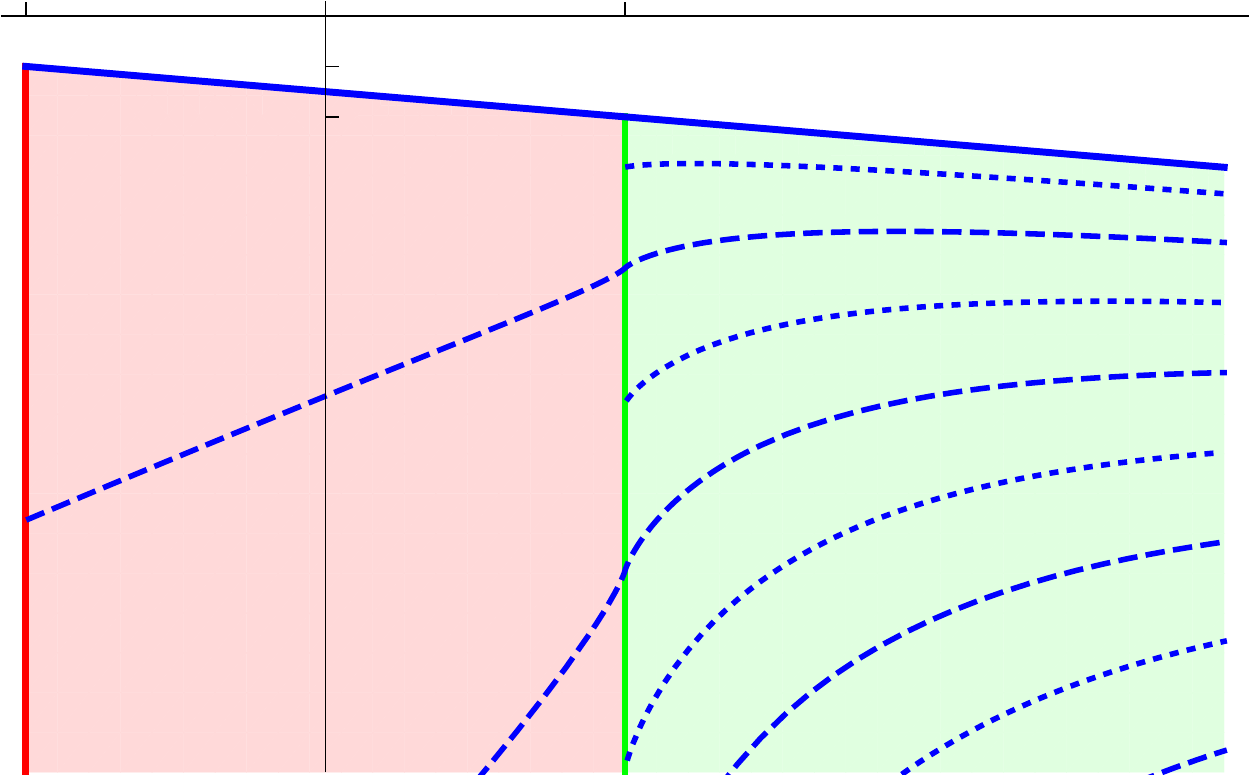}}
\put(45,14){\includegraphics[width = 0.4\textwidth]{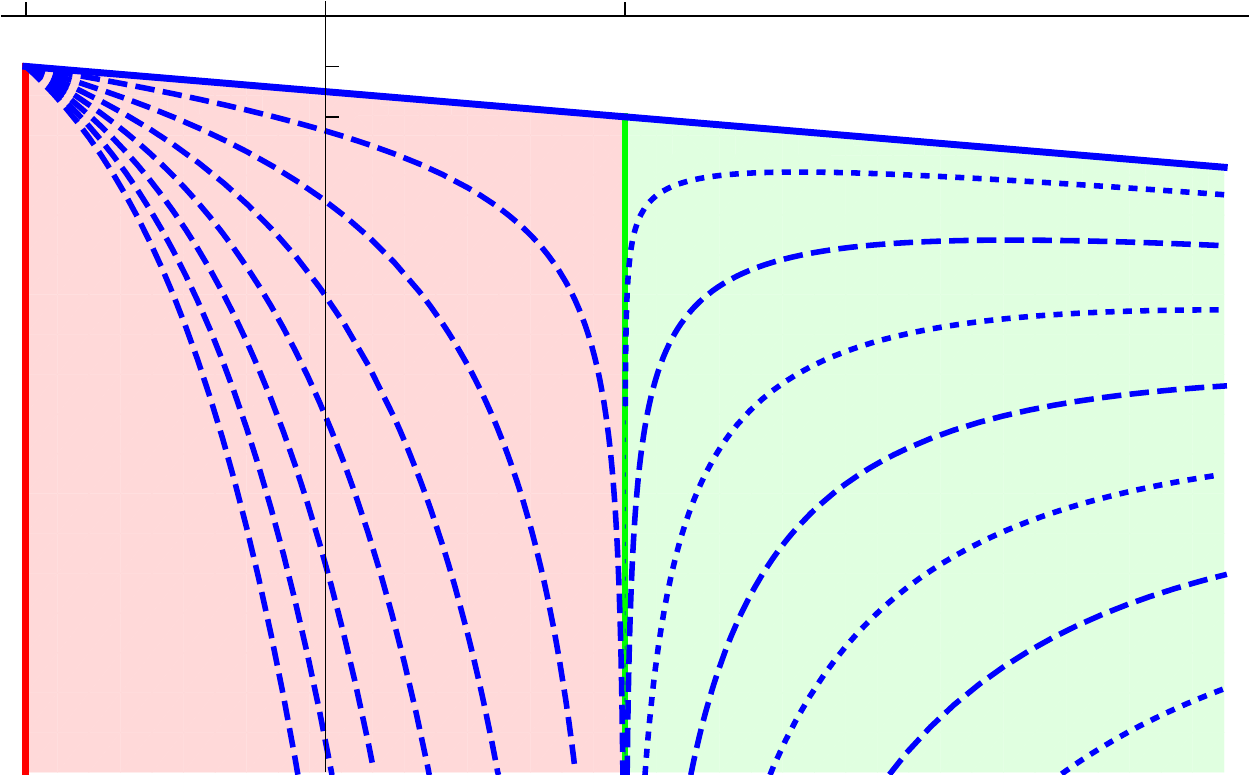}}
\put(10.5,12){static counterparts}
\put(44.5,12){translationally invariant counterparts}
\put(40,38.75){$E$}
\put(-0.75,39.5){$-\mu^2$}
\put(20,39.5){$\mu^2$}
\put(9.25,40){$\wp \left( a \right)$}
\put(39.75,36.75){$\underline{n_1}$}
\put(40.5,33.75){\bf{2}}
\put(40.5,31.75){$3$}
\put(40.5,29.75){$4$}
\put(40.5,27.75){$5$}
\put(40.5,25.75){$6$}
\put(40.5,23.25){$7$}
\put(40.5,20.5){$8$}
\put(40.5,17.5){$9$}
\put(40.5,14){$10$}
\put(84,38.75){$E$}
\put(43.25,39.5){$-\mu^2$}
\put(64,39.5){$\mu^2$}
\put(53,40){$\wp \left( a \right)$}
\put(83.75,36.75){$\underline{n_0}$}
\put(84.5,33.75){\bf{0}}
\put(84.5,31.75){$1$}
\put(84.5,29.75){$2$}
\put(84.5,27.5){$3$}
\put(84.5,25.25){$4$}
\put(84.5,22.5){$5$}
\put(84.5,19.5){$6$}
\put(84.5,16){$7$}
\put(6,0){\includegraphics[height = 0.1\textwidth]{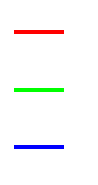}}
\put(11,1){GKP limit/oscillating hoops}
\put(11,4.25){giant magnons/single spikes}
\put(11,7.25){hoop/BMN partiple}
\put(46,0){\includegraphics[height = 0.095\textwidth]{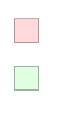}}
\put(51,2.5){rotating counterparts}
\put(51,6.5){oscillating counterparts}
\put(6,0){\line(0,1){10}}
\put(6,0){\line(1,0){76}}
\put(82,0){\line(0,1){10}}
\put(6,10){\line(1,0){76}}
\end{picture}
\end{center}
\vspace{-10pt}
\caption{The moduli space of elliptic string solutions}
\vspace{5pt}
\label{fig:elliptic_string_moduli}
\end{figure}

The elliptic string solutions are a two-parameter family of solutions, in our language being the parameters $E$ and $a$. The advantage of our parametrization is that only one of the two parameters (the integration constant $E$) affects the corresponding solution of the Pohlmeyer reduced system. The worldsheets of the solutions being characterized by the same constant $E$ comprise an associate (Bonnet) family \cite{Pastras:2016vqu}. Demanding appropriate periodicity conditions, restricts one of the two parameters to be discrete, or in other words the moduli space of the elliptic string solutions with appropriate periodicity conditions is a discretely infinite set of one-dimensional curves. Figure \ref{fig:elliptic_string_moduli} depicts the moduli space of elliptic string solutions and visualises their classification according to their Pohlmeyer counterpart.

\setcounter{equation}{0}
\section{Energy and Angular Momentum}
\label{sec:dispersion}
The $\mathbb{R}^t \times $S$^2$ target space has the symmetry of time translations, leading to a conserved energy and that of SO(3) rotations, leading to a conserved angular momentum.

Considering solutions with appropriate periodic conditions, the string energy is given by
\begin{equation}
E_{0/1} = \left| \frac{\delta L}{\delta \partial_0t} \right| =  T \int_0^{{n_{0/1} \delta \sigma_{0/1}}} {\frac{{\partial t_{0/1}}}{{\partial \sigma^0 }}d\sigma^1 }  = T {{n_{0/1}R\mu \delta \sigma_{0/1}}} = \frac{2 T n_{0/1}R{\mu ^2}{\omega _1}}{\sqrt {{x_{3/2}} - \wp \left( a \right)}}  , 
\label{eq:dispersion_energy}
\end{equation}
where $n_{0/1} \in \mathbb{Z}$ when $E>\mu^2$ and $n_{0/1} \in 2\mathbb{Z}$ when $E<\mu^2$. The above expression is indeterminate in the GKP limit of solutions with translationally invariant counterparts ($\wp (a) = x_3$, $n_0 = 0$). In this case, the energy assumes the value $E_0 = 2 \pi T R \mu / \sqrt{x_1 - x_3}$.

Similarly, the $z$-component of the angular momentum is given by
\begin{equation}
\begin{split}
{J_{0/1}} &= \frac{\delta L}{\delta \partial_0\varphi} = T R^2 \int_0^{n_{0/1}\delta {\sigma _{0/1}}} {{{\sin }^2}\theta_{0/1} \frac{{\partial \varphi_{0/1} }}{{\partial \sigma^0 }}d\sigma^1 } \\
&= \mp \frac{T R^2}{\mu }\sqrt {\frac{{{x_{2/3}} - \wp \left( a \right)}}{{{x_1} - \wp \left( a \right)}}} \int_0^{n_{0/1}\delta {\sigma _{0/1}}} {\left( {\wp \left( {\gamma \left( {{\sigma ^{0/1}} - \beta {\sigma ^{1/0}}} \right) + {\omega _2}} \right) - {x_{2/3}}} \right)d\sigma^1 } \\
&=  \pm \frac{{2 T n_{0/1} R^2 \left( {\zeta \left( {{\omega _1}} \right) + {x_{2/3}}{\omega _1}} \right)}}{{\sqrt {{x_1} - \wp \left( a \right)} }} .
\end{split}
\label{eq:dispersion_angular_momentum}
\end{equation}

In the following, we define $\mathcal{E}_{0/1} := E_{0/1} / (2TR)$ and $\mathcal{J}_{0/1} := J_{0/1} / (2TR^2)$. The mismatch of the $R$ factors in these definitions is due to the fact that we have considered time as an independent dimension not related to the radius of the sphere. Had we considered $\mathbb{R}^t \times$S$^2$ as a submanifold of an AdS$_n \times$S$^n$ space with a dual boundary description, the time would have been part of the AdS$_n$, which has the same radius as that of the sphere, effectively measuring time in units of $R$. We also recall that the angular opening $\delta \varphi$, which is associated to the quasi-momentum in the dual theory is given by
\begin{equation}
\delta \varphi_{0/1} = \mp 2 \omega_1 \left( {i\zeta \left( {{\omega _1}} \right)\frac{a}{\omega_1} - \left( {i\zeta \left( a \right) + \sqrt {\frac{{\left( {{x_1} - \wp \left( a \right)} \right)\left( {{x_{2/3}} - \wp \left( a \right)} \right)}}{{{x_{3/2}} - \wp \left( a \right)}}} } \right)} \right).
\end{equation}

In the positive double root limit, the Weierstrass functions degenerate to simple trigonometric functions. It is a matter of algebra to show that in this limit and in the case of static counterparts, the energy and angular momentum diverge, due to the divergence of $\omega_1$ and it holds that
\begin{align}
{\mathcal{E}_0} + \frac{{\delta {\varphi _0}}}{2} &=  - 2i\mu a =  - \arcsin \mathcal{J} ,\\
\mathcal{E}_1 - \mathcal{J}_1 &= n_1 \sin \left( - i \mu a \right) = n_1 \sin \frac{\delta \varphi_1}{2} ,
\end{align}
which is the very well known dispersion relations of the single spikes and giant magnons.

In this parametrization, it is also simple to study the limit of the speeding strings. As $a / \omega_2 \to 0$ the angular opening $\delta\varphi$ tends to zero. whereas the energy and angular momentum remain finite. In this limit, the angular opening, energy and angular momentum assume the form
\begin{align}
\delta \varphi_{0/1} &\simeq \mp 2 \left( \zeta \left( \omega_1 \right) + x_{3/2} \omega_1 \right) \left( i a \right) + \mathcal{O} \left( a^3 \right) ,\\
\mathcal{E}_{0/1} &\simeq n_{0/1} \mu^2 \omega_1 \left( i a \right) + \mathcal{O} \left( a^3 \right) ,\\
\mathcal{J}_{0/1} &\simeq \pm n_{0/1} \left( \zeta \left( \omega_1 \right) + x_{2/3} \omega_1 \right) \left( i a \right) + \mathcal{O} \left( a^3 \right) ,
\end{align}
implying that
\begin{equation}
\mathcal{E}_{0/1} - \mathcal{J}_{0/1} \simeq \frac{1}{2} n_{0/1} \delta \varphi .
\end{equation}
This is compatible to the giant magnon case since in this limit $\delta \varphi \to 0$.

The expressions \eqref{eq:dispersion_energy} and \eqref{eq:dispersion_angular_momentum} that provide the energy and angular momentum of the string in terms of the Weierstrass functions can be used to convert the problem of the specification of the dispersion relation to an algebraic problem with the help of appropriate properties of the latter functions. For example, let us consider the special case $a = - \omega_2 / 2$. This is a one-dimensional family of solutions, which in the case of static counterparts, contains the giant magnon with angular opening equal to $\pi / 2$. The Weierstrass functions obey the following quarter period relations
\begin{equation}
\wp \left( {\frac{{{\omega _2}}}{2}} \right) = {e_3} - \sqrt {\left( {{e_3} - {e_1}} \right)\left( {{e_3} - {e_2}} \right)}  =  - \frac{E}{6} - \frac{{{\mu ^2}}}{2} - \mu \sqrt {\frac{{E + {\mu ^2}}}{2}} 
\end{equation}
and
\begin{equation}
\begin{split}
\zeta \left( {\frac{{{\omega _2}}}{2}} \right) &= \frac{1}{2}\left( {\zeta \left( {{\omega _2}} \right) - i\sqrt {2\sqrt {\left( {{e_3} - {e_1}} \right)\left( {{e_3} - {e_2}} \right)}  - 3{e_3}} } \right) \\
&= \frac{1}{2}\left( {\zeta \left( {{\omega _2}} \right) - i\left( {\sqrt {\frac{{E + {\mu ^2}}}{2}}  + \mu } \right)} \right) .
\end{split}
\end{equation}
Using the above properties, the angular opening of the string assumes the form
\begin{equation}
\delta {\varphi _{0/1}}\left( {E, - \frac{{{\omega _2}}}{2}} \right) =  \pm \left( { - \frac{\pi }{2} + {\omega _1}\left( {\sqrt {\frac{{E + {\mu ^2}}}{2}}  \pm \mu } \right)} \right) ,
\end{equation}
whereas the energy of the string is written as
\begin{align}
{\mathcal{E}_0}\left( {E, - \frac{{{\omega _2}}}{2}} \right) &= \mu {\omega _1}{\left( {\frac{{E + {\mu ^2}}}{{2{\mu ^2}}}} \right)^{ - \frac{1}{4}}},\\
{\mathcal{E}_1}\left( {E, - \frac{{{\omega _2}}}{2}} \right) &= \mu {\omega _1}{\left( {{{\left( {\frac{{E + {\mu ^2}}}{{2{\mu ^2}}}} \right)}^{\frac{1}{2}}} + 1} \right)^{ - \frac{1}{2}}} .
\end{align}
This implies that the integration constant $E$ can be expressed as the algebraic function of the quantity $({{\delta {\varphi _{0/1}} \pm \pi /2}})/{{{\mathcal{E}_{0/1}}}}$, which solves the equation,
\begin{align}
\frac{{\delta {\varphi _0} + \pi /2}}{{{\mathcal{E}_0}}} &= {\left( {\frac{{E + {\mu ^2}}}{{2{\mu ^2}}}} \right)^{\frac{3}{4}}} + {\left( {\frac{{E + {\mu ^2}}}{{2{\mu ^2}}}} \right)^{\frac{1}{4}}} ,\\
\frac{{\delta {\varphi _1} - \pi /2}}{{{\mathcal{E}_1}}} &= \left( {1 - {{\left( {\frac{{E + {\mu ^2}}}{{2{\mu ^2}}}} \right)}^{\frac{1}{2}}}} \right){\left( {1 + {{\left( {\frac{{E + {\mu ^2}}}{{2{\mu ^2}}}} \right)}^{\frac{1}{2}}}} \right)^{\frac{1}{2}}} .
\end{align}
These equations are equivalent to cubic equations for $E / \mu^2$. Once this function is specified, it can be substituted in the expression \eqref{eq:dispersion_angular_momentum} in order to acquire an analytic dispersion relation connecting $\mathcal{E}$, $\mathcal{J}$ and $\delta \varphi$ that characterises the string solutions with $a = - \omega_2 / 2$, arbitrarily far from the infinite size limit.  Notice that the real period $\omega_1$ can be expressed as an algebraic function of $\mathcal{E}$ and $\delta \varphi$ through equation \eqref{eq:dispersion_energy}. So the only transcendental part of the dependence of the angular momentum on $\mathcal{E}$ and $\delta \varphi$ is through $\zeta (\omega_1)$ or equivalently the complete elliptic integral of the second kind, which is finite everywhere in $\left[0,1\right]$.

This procedure can be generalized. Consider the more general case $a = - 2 q \omega_2$, $q \in \mathbb{Q}$. This is a one-dimensional sector of the moduli space, which, in the case of static counterparts, contains a giant magnon solution obeying appropriate periodicity conditions with $\delta \varphi = 2 q \pi$ (of course this is going to have self-intersections unless $q$ is of the form $1/n$, $n \in \mathbb{Z}$). The functions $\wp \left( 2 m z / n \right)$ with $m,n \in \mathbb{Z}$ and $\wp \left( z \right)$ are both elliptic functions with periods $2 n \omega_1$ and $2 n \omega_2$. Therefore they are algebraically related. The above argument for $z = \omega_2$ implies that $\wp \left( 2 q \omega_2 \right)$ is an algebraic function of the root $e_3$. 

Furthermore, the Weierstrass zeta function obeys
\begin{equation}
\zeta \left( {z + w} \right) = \zeta \left( z \right) + \zeta \left( w \right) + \frac{1}{2}\frac{{\wp '\left( z \right) - \wp '\left( w \right)}}{{\wp \left( z \right) - \wp \left( w \right)}},\quad \zeta \left( {2z} \right) = 2\zeta \left( z \right) + \frac{{\wp ''\left( z \right)}}{{2\wp '\left( z \right)}}.
\end{equation}
As a result of the Weierstrass equation ${\left( {\wp '} \right)^2} = 4{\wp ^3} - {g_2}\wp  - {g_3}$ and its derivative $\wp '' = 6{\wp ^2} - {g_2}/2$, $\wp' \left( z \right)$ and $\wp'' \left( z \right)$ are algebraic functions of $\wp \left( z \right)$.
Iterative use of the above formulas implies that $\zeta \left( n z \right) = n \zeta \left( z \right) + f_n \left( \wp \left( z \right) \right)$, where $f_n$ is an algebraic function. Applying the above for $z = 2 m \omega_2 / n$ results in the zeta Weierstrass function $\zeta \left( 2 m \omega_2 / n \right)$ being equal to $ 2 m \zeta \left( \omega_2  \right) / n$ plus an algebraic function of the root $e_3$, or equivalently an algebraic function of the ratio $E / \mu^2$, i.e.
\begin{equation}
\zeta \left( {{{{2 q \omega _2}}}} \right) = {{2 q \zeta \left( {{\omega _2}} \right)}} + {f_q}\left( {{E / \mu^2}} \right),
\end{equation}
The specification of these algebraic functions may be a difficult task in practice. Indicatively in the case $q = 1 / 3$, $\wp \left( 2 \omega_2 / 3 \right)$ is the smallest root of the quartic equation $48 P^4 - 24 g_2 P^2 -48 g_3 P -g_2^2 = 0$ and $\zeta \left( 2 \omega_2 / 3 \right) = {{2 \zeta \left( {{\omega _2}} \right)}}/{3} - \left( \wp \left( 2 \omega_2 / 3 \right) \right)^{1/2}$.

Once these functions have been specified, the angular opening and the energy of the string assume the form
\begin{align}
\delta {\varphi _{0/1}}\left( {E, - {{{2 q \omega _2}}}} \right) &= \pm \left( { - q {\pi } + {\mu \omega _1}{g_q}\left( {{E}/{{{\mu ^2}}}} \right)} \right) , \label{eq:dispersion_trajectory_df}\\
{\mathcal{E}_{0/1}}\left( {E, - {{{2 q \omega _2}}}} \right)& = {\mu \omega _1}{h_q}\left( {{E}/{{{\mu ^2}}}} \right) ,
\end{align}
where ${g_q}\left( {{E}/{{{\mu ^2}}}} \right)$ and ${h_q}\left( {{E}/{{{\mu ^2}}}} \right)$ are algebraic functions of $E / \mu^2$. Therefore, the ratio $E / \mu^2$ is an algebraic function of the quantity ${{{\left( {\delta {\varphi _0} \pm q \pi} \right)}}/{{{\mathcal{E}_{0/1}}}}}$, i.e.
\begin{equation}
E = {\mu ^2}{F_q}\left( {\frac{{ {\delta {\varphi _{0/1}} \pm q \pi} }}{{{\mathcal{E}_{0/1}}}}} \right) .
\end{equation}
Once this algebraic function is specified, it can be substituted in \eqref{eq:dispersion_angular_momentum} to provide a closed formula for the dispersion relation of elliptic strings that satisfy $a = - 2 q \omega_2 $. 

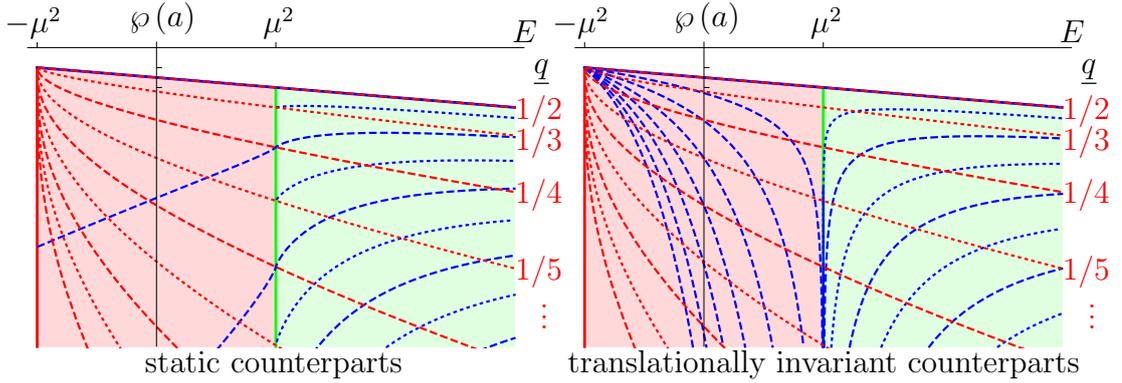
\begin{figure}[ht]
\vspace{10pt}
\begin{center}
\begin{picture}(88,32.5)
\put(1,4){\includegraphics[width = 0.4\textwidth]{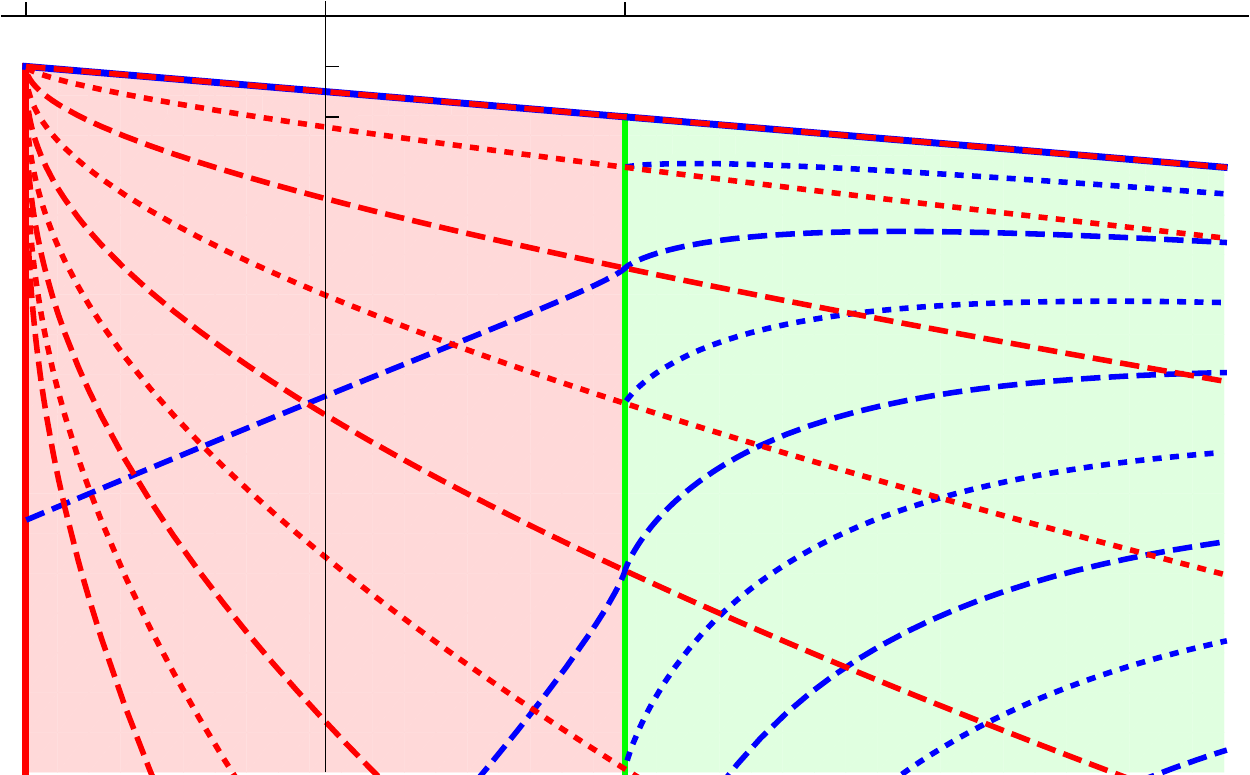}}
\put(45,4){\includegraphics[width = 0.4\textwidth]{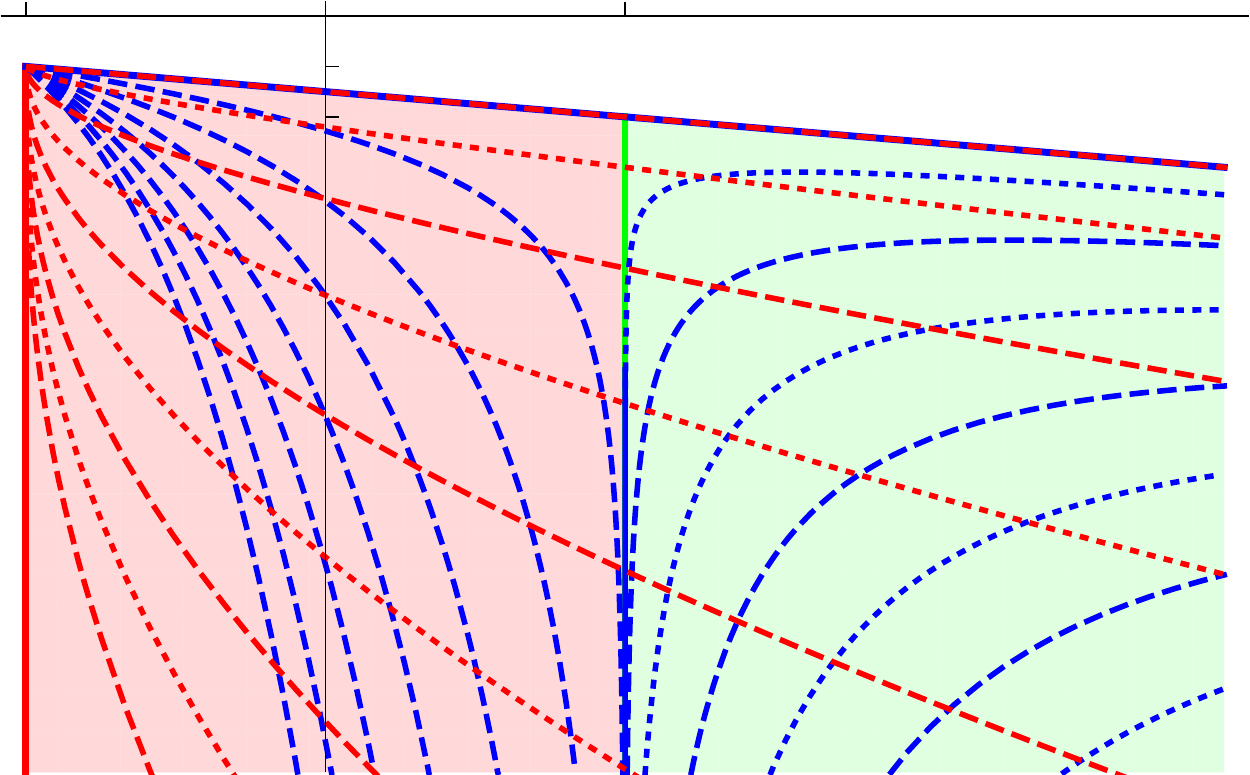}}
\put(10.5,2){static counterparts}
\put(44.5,2){translationally invariant counterparts}
\put(40,28.75){$E$}
\put(-0.75,29.5){$-\mu^2$}
\put(20,29.5){$\mu^2$}
\put(9.25,30){$\wp \left( a \right)$}
\put(41.75,26.25){$\underline{q}$}
\put(40.25,22.5){\textcolor{red}{1/2}}
\put(40.25,20){\textcolor{red}{1/3}}
\put(40.25,15.75){\textcolor{red}{1/4}}
\put(40.25,9.5){\textcolor{red}{1/5}}
\put(42.25,5.5){\textcolor{red}{$\vdots$}}
\put(84,28.75){$E$}
\put(43.25,29.5){$-\mu^2$}
\put(64,29.5){$\mu^2$}
\put(53,30){$\wp \left( a \right)$}
\put(85.75,26.25){$\underline{q}$}
\put(84.25,22.5){\textcolor{red}{1/2}}
\put(84.25,20){\textcolor{red}{1/3}}
\put(84.25,15.75){\textcolor{red}{1/4}}
\put(84.25,9.5){\textcolor{red}{1/5}}
\put(86.25,5.5){\textcolor{red}{$\vdots$}}
\end{picture}
\end{center}
\vspace{-10pt}
\caption{The trajectories in the moduli space where the dispersion relation can be specified analytically}
\vspace{5pt}
\label{fig:moduli_trajectories}
\end{figure}

Since the set of rational numbers is a dense subset of the real numbers, the union of the trajectories $a = - 2 q \omega_2 $, where the dispersion relation assumes an analytic form, is a dense subset of the moduli space of the elliptic string solutions. Figure \ref{fig:moduli_trajectories} shows how the $a = - 2 q \omega_2 $ trajectories lie in the moduli space.

The above process cannot be applied in the case of the GKP limit, i.e. the specific selection $q=1/2$. In this case, the angular opening is not a function of the integration constant $E$, but it simply equals $\delta \varphi_1 = \pi$, i.e. the algebraic function $g_q$ in equation \eqref{eq:dispersion_trajectory_df} vanishes. Therefore, the integration constant $E$ cannot be specified algebraically by an appropriate linear combination of the energy and the angular opening, but it requires the inversion of the elliptic integral that relates it to the string energy. This cannot be performed analytically; usually this inversion is performed perturbatively around the infinite size limit \cite{Arutyunov:2006gs,Ahn:2008sk,Floratos:2013cia,Floratos:2014gqa}.

\setcounter{equation}{0}
\section{Discussion}
\label{sec:discussion}

We applied a systematic approach to the construction of classical string solutions on $\mathbb{R}^t \times $S$^2$. Using a specific class of solutions of the Pohlmeyer reduced theory, i.e. the sine-Gordon equation, which are expressed in terms of elliptic functions, we were able to develop a unified description of all known genus one string solutions on $\mathbb{R}^t \times $S$^2$. Our approach is based on a clever choice of the worldsheet parametrization that leads to equations of motion for the classical string, which are solvable via separation of variables.

The fact that our method can be applied successfully, reproducing all known genus one solutions and providing a unified framework is not accidental. The NLSM is integrable, and, thus, it can be solved using finite gap integration. It is known that any smooth one-gap potential is equivalent to an appropriate $n=1$ \Lame potential \cite{Hochstadt:1965}. Thus, in the case of elliptic solutions, the equations of motion are in principle reducible to the $n=1$ \Lame problem. This is precisely what it is achieved via the application of the Pohlmeyer reduction inversion technique. Since the spiky strings and their various special limits are the most general genus one classical string solutions \cite{Vicedo:2007rp,Vicedo:2011zz}, our approach achieves the inversion of the Pohlmeyer reduction and it is equivalent to the finite gap integration in the case of genus one. 

An advantage of our unified description is the convenience in studying and comparing the properties of the string solutions to those of their Pohlmeyer counterparts. For example, rigidly rotating strings have counterparts which can be set static after an appropriate worldsheet boost. On the other hand, wave propagating solutions have counterparts that can be set translationally invariant after an appropriate worldsheet boost. Spikes occur at points where the Pohlmeyer field assumes a value equal to an integer multiple of $2 \pi$. These points are always moving at the speed of light. Finally, the topological charge in the sine-Gordon theory is mapped to the number of spikes of the string. This mapping of properties provides a nice geometric picture to the Pohlmeyer reduction and enhances our intuition on the dynamics of string propagation on $\mathbb{R}^t \times $S$^2$. Table \ref{tb:correspondence} summarizes the mapping of the properties of the strings to those of their Pohlemeyer counterparts.
\begin{table}[ht]
\centering
\begin{tabular}{|p{0.4\textwidth}|p{0.4\textwidth}|}
\hline
NLSM  & sine-Gordon \\
\hline\hline
two-parameter family of solutions & only one of the two parameters affects the solution \\
\hline
angular frequency \newline extremal altitudes & gauge in which the solution is either static or translationally invariant\\
\hline
degenerate one-dimensional worldsheet (BMN particle, hoops)& vacuum solution\\
\hline
strings asymptotically reaching the equator (giant magnons, single spikes) & kink or instanton solutions \\
\hline
rigid rotation/wave propagation & static/translationally invariant solutions (at some frame) \\
\hline
spike & $\varphi = 2 n \pi , \quad n \in \mathbb{Z}$\\
\hline
number of spikes & topological charge \\
\hline
spiky strings/non-spiky strings or strings with equal number of spikes and anti-spikes & rotating/oscillating solutions --- \newline multi fermion states/bosonic condensate states in the dual Thirring model \\
\hline
\end{tabular}
\caption{A dictionary between the NLSM and the sine-Gordon model}
\label{tb:correspondence}
\end{table}

The Weierstrass elliptic function is the natural parametrization for the study of genus one solutions, since it uniformizes the torus. The manifestation of the latter is the simple unified description of this class of classical string solutions in terms of the ``effective energy'' $E$ of the sine-Gordon reduced system and the purely imaginary parameter $a$. Adopting this parametrization significantly simplifies the expressions for the conserved charges of the string and facilitates the study of the corresponding dispersion relation. In particular, we identify a set of one-dimensional trajectories in the moduli space, where it is possible to express the moduli as an algebraic function of the ratio of the energy and the angular opening, allowing the expression of the dispersion relation in a closed form, arbitrarily far away from the infinite size limit. These trajectories compose a dense subset of the moduli space.

Another interesting feature that emerges from the properties of the sine-Gordon equation has to do with its well known duality with the Thirring model. The topological charge of the sine-Gordon theory is mapped to the fermion number in the Thirring model. Therefore, the number of spikes has a naive interpretation as a fermion number. In this picture, the strings with rotating Pohlmeyer counterparts have the natural interpretation of fermionic objects of the theory, whereas the strings with oscillating counterparts have the interpretation of bosonic condensates of the latter. The study of elliptic strings in this context would have an enhanced interest in view of the S-duality of the type IIB superstring theory in AdS$_n \times$S$^n$ spaces. In this case, such elliptic solutions could provide a quantitative tool to understand the role of the sine-Gordon/Thirring duality as S-duality in the Pohlmeyer reduced theory.

The presented techniques can be directly generalized to higher dimensional spheres and to AdS$_n\times$S$^n$ spaces. As long as S$^n$ is concerned, when $n$ is even, the eigenvalues of the problem will have the same structure as in the presented S$^2$ case: there will be an odd number of enhanced space embedding functions, which will be organised in several pairs, each being associated to a positive eigenvalue connected to a Bloch wave eigenstate of the associated $n=1$ \Lame problem and a single one that will be associated with a vanishing eigenvalue, and, thus, connected to an eigenstate of the $n=1$ \Lame problem lying at the margin of a band. When $n$ is odd, there will be an even number of enhanced space coordinates, which will be simply organised in pairs each associated with a positive eigenvalue. Such solutions have been constructed with other methods in the literature \cite{Kruczenski:2006pk}. Further extending to AdS$_n\times$S$^n$, which is of particular interest towards holographic applications, requires the combination of the results presented in this work with those of \cite{bakas_pastras}. The elliptic strings on AdS spaces form some qualitatively distinct classes due to the form of the metric in the enhanced space (which is $\mathbb{R}^{(2,n-1)}$). It would be interesting to study how these classes get combined with the elliptic strings on the sphere and how they differ in terms of their dispersion relation or other geometric characteristics.

\subsection*{Acknowledgements}
The research of G.P. is funded by the ``Post-doctoral researchers support'' action of the operational programme ``human resources development, education and long life learning, 2014-2020'', with priority axes 6, 8 and 9, implemented by the Greek State Scholarship Foundation and co-funded by the European Social Fund - ESF and National Resources of Greece.

The authors would like to thank M. Axenides and E. Floratos for useful discussions.

\end{document}